\documentclass[%reprint,
superscriptaddress,
 amsmath,amssymb,
 aps,pre,hyphens,floatfix
twocolumn]{revtex4-2}
\usepackage[running]{lineno}
\usepackage[]{amsmath}
\usepackage{mathrsfs}
\usepackage{mathtools} 
\usepackage{enumerate}
\usepackage[usenames,dvipsnames,svgnames,table]{xcolor}
\usepackage{IEEEtrantools}
\usepackage{graphicx}% Include figure files
\usepackage{array}
\usepackage{multirow}
\usepackage{verbatim}
\usepackage{soul}
\usepackage[normalem]{ulem}
\usepackage{tabularx}
\usepackage{booktabs,siunitx}
\usepackage[T1]{fontenc}
\usepackage{array, xcolor}

\usepackage[hidelinks]{hyperref} 
\hypersetup{
    colorlinks,
    linkcolor={blue!50!black},
    citecolor={blue!50!black},
    urlcolor={blue!80!black}
}
\usepackage[hyphenbreaks]{breakurl}

\usepackage{csquotes}

\newcommand{\red}[1]{\textcolor{black}{#1}}

\newcommand{\mycomment}[1]{}
\usepackage{nameref}
\usepackage{titlesec}

\newcommand{\av}[1]{\langle {#1} \rangle}
\newcommand{\stk}[1]{\ifmmode\text{\sout{\ensuremath{#1}}}\else\sout{#1}\fi}

\begin{document}
\title{Fluctuation-dissipation of the Kuramoto model on fruit-fly connectomes}

\author{G\'{e}za \'{O}dor}
\email[]{odor.geza@ek.hun-ren.hu}
\affiliation{Institute of Technical Physics and Materials Science, Center for Energy Research, P.O. Box 49, H-1525 Budapest, Hungary}
\author{Istv\'an Papp}
\affiliation{Institute of Technical Physics and Materials Science, Center for Energy Research, P.O. Box 49, H-1525 Budapest, Hungary}
\author{Gustavo Deco}
\affiliation{Center for Brain and Cognition, Theoretical and Computational Group,
Universitat Pompeu Fabra / ICREA, Barcelona, Spain}

\begin{abstract}
We investigate the distance from equilibrium using the Kuramoto model via the degree of fluctuation-dissipation violation as the consequence 
of different levels of edge weight anisotropies. This is achieved by solving the synchronization equations on the raw, homeostatic weighted 
and a random inhibitory edge variant of a real full fly (FF) connectome, containing $\simeq 10^5$ neuron cell nodes. 
We investigate these systems close to their synchronization transition critical points. While the topological(graph) 
dimension is high: $d \simeq 6$ the spectral dimensions of the variants, relevant in describing the synchronization behavior, 
are lower than the upper critical dimension: $d_s \simeq 2 < d_c=5$, suggesting relevant fluctuation effects and non 
mean-field scaling behavior. 
By measuring the auto-correlations and the auto-response functions for small perturbations we calculate the 
fluctuation-dissipation ratios (FDR) for the different variants of different anisotropy levels of the FF connectome. 
Numerical evidence is presented that the FDRs follow the level of anisotropy of these non-equilibrium systems in agreement with the expectations, with increasing temporal oscillations. 
Numerical values for the aging exponents provide confirmation for the predictions of the 
generalized time-translational-invariance theory.
We compare the non-reciprocal connectome results with those on a symmetric Erd\H os-R\'enyi random graph of similar size and provide estimates for the Kuramoto model aging behavior.
We also present a network analysis of the FF connectome and calculate the level of hierarchy, also related to 
the anisotropy. Finally, we provide some partial results for the periodically forced Shinomoto-Kuramoto model 
describing fluctuations in the non-resting state. We provide numerical evidence that fluctuations are maximal in the resting state of the critical brain.
\end{abstract}

\flushbottom
\maketitle

\thispagestyle{empty}

\section{Introduction}

Thanks to the explosion of informatics and precise neurodata analysis large structural 
connectomes are available in different repositories 
(see for example: ~\cite{AllenReferenceAtlas2011,down-hemibrain1.0.1}) 
by now, but one has to be cautious with non-proofread ones~\cite{CIRUNAY2026118425}.  
Here we used the fruit-fly connectome as a complex, empirical case study of a directed graph, 
as close to the reality of a biological neural system as possible, rather than as 
the primary driver of the research.
In this study, we harness this modeling framework to introduce perturbability metrics, which assess how external
perturbations influence brain dynamics. Specifically, perturbability quantifies the brain’s susceptibility to external inputs, 
while perturbational recovery measures the system’s capacity to return from a pathological state toward a healthy dynamical regime.

Crucially, the theoretical backbone of these measures is grounded in the Fluctuation-Dissipation Theorem (FDT) from statistical physics~\cite{FDT}. 
Neural systems, being inherently out of equilibrium, naturally exhibit complex fluctuations across spatial and temporal scales. 
The FDT characterizes the fundamental relationship between these spontaneous fluctuations and the system’s response to external 
perturbations. In the brain, this reflects a balance between internal noise and dissipation mechanisms. 
Deviations from FDT predictions highlight a departure from equilibrium, often marking an increased sensitivity or altered responsiveness 
to stimuli—a hallmark of pathological brain states.

Non-equilibrium systems can be characterized by the violation of the fluctuation-dissipation (vFD) relation~\cite{FD-Cug,MARCONI2008111}).
It is related to the non-equilibrium nature of the dynamics; the prominent example is the 
case of systems with absorbing states~\cite{odor04}. 
Spatial anisotropy also breaks the detailed balance condition
(necessary to be in equilibrium state). 
A prominent example is when we add an up/down symmetry-breaking term to surface growth, 
which changes the equilibrium Edwards Wilkinson growth to the non-equilibrium KPZ 
equation~\cite{odor04}. Similarly, when we add spatial anisotropy to the diffusion 
or reaction terms, like in the case of driven diffusive particle models, vFD is achieved. 
Modular structure, or other constraints in the dynamics, may also generate vFD.

Very recently the vFD in neural systems has been studied by~\cite{DecoPhysRevE.108.064410,Monti2025PRR} and the relation 
between vFD and the asymmetry of the interactions has been demonstrated using empirical human neuroimaging data as well as 
whole-brain models. Using this approach they uncovered the thermodynamic hierarchy of conscious states and quantified 
disruptions in disorders of consciousness. They demonstrated that wakefulness shows significant violations of FDT, 
whereas deep sleep exhibits a major decrease in FDT violations, indicating a flattening of the brain's 
hierarchical organization toward more symmetrical interactions. Evidence was found that global and resting-state 
network dynamics in patients with disorders of consciousness is closer to equilibrium than in healthy controls, 
decreasing stepwise with lower consciousness.

In this work we extend this research, by taking into account the critical 
brain hypothesis~\cite{ChialvBak1999_LearningMistakes,Chi10,Beggs12}, i.e we investigate the vFD on large scale 
whole brain models tuned at the critical synchronization point. 
This is based on our previous whole brain Kuramoto model studies of different structural connectomes, 
such as the fruit-fly hemi-brain (HB) and large human gray mater graphs~\cite{KKIdeco,CCrev,Flycikk,sniccikk}.
We investigate here what consequences of this hypothesis implies for the vFD results.

Note, that for dynamical simulations we could have used large synthetic complex networks, with randomness 
as we had done before, like Erd\H os R\'enyi graphs~\cite{Kurcikk}, Watts-Strogatz and Benjamini-Berger type long-link graphs~\cite{Kurcikk,KurCC} or hierarchical modular ones~\cite{HMNcikk,HPTcikk}, 
but these assume some features: dimensions, symmetries, modularity features... etc, relevant for 
the critical behavior. In order to extend the previous studies towards more brain like systems
one should tuned them manually, before the simulations. Instead of that, now we decided to 
use a real FF connectome, which sets these relevant features. To be sure we present the necessary 
structural graph analysis in the first part of this publication and later we test by simulations what happens with the dynamical behavior if vary the level of interaction anisotropy, i.e. the reciprocity symmetry of edges.
In this work we investigate the relation of anisotropy and non-equilibrium critical dynamical
behavior within the framework of this connectome, using an oscillatory model for the function. 
We studied the network for three different anisotropy levels: the original network (o); the case with 
renormalized weights (w); and the case with $20\%$ negative weights chosen randomly to model inhibition (i), 
see section \enquote{\nameref{sec:FFaVAR}}.

Far-from-equilibrium systems often display dynamical scaling, even if the stationary state is very far from being 
critical~\cite{henkel2011non,Henk-age-conf}. 
Criticality, often occurring at continuous, second-order phase transitions, is common in nature and beneficial for systems. 
In neural systems, criticality allows for the generation of working memory, spontaneous long-range interactions, and maximal 
sensitivity to external signals due to diverging correlations and fluctuations~\cite{COCCHI2017132,MArep,CCrev}. 
Information-processing is also optimal near the critical point~\cite{KC,Chi10,Larr,MArep}, leading systems to self-organize 
close to criticality (SOC)~\cite{SOC,Chi10}, or slightly below it to avoid over-excitation. 
This can also happen in the dynamical, semi-critical so called Griffiths phase~\cite{Griffiths,Munoz2010,MM,HMNcikk,OdorPRE2016}.

Another important, open question in the level of hierarchical anisotropy of neural systems.
The modular structure of brain is well known, but whether they follow hierarchical or
non-hierarchical organization is debated. To fill this gap we determined the modules of
the largest exactly known brain, the full fruit fly connectome by community detecting algorithm 
and run an up to date anisotropy measuring method, based on directed random walk approach.

Now we summarize in nutshell the FDT through the relation between the response and the auto-correlation functions. 
The auto-response function $\mathcal{R}(t,s)$ of an observable $E(t)$ of the system to a small enough 
perturbation $h$ at time $s$ is given by
\begin{equation}\label{eq:R}
    \mathcal{R}(t,s) = \frac{\delta E_h(t,s)} {\delta h} |_h \to 0 \ ,
\end{equation}
whereas the auto-correlation between the start $s$ and measurement $t$ times is 
defined as
\begin{equation}\label{eq:Adef}
 A(t,s) = \langle E(t) E(s)\rangle,
\end{equation}
\noindent here $\langle \cdot \rangle$ states for the ensemble average.
At the critical point $A(s,t)$ and $\mathcal{R}(s,t)$ are expected to decay asymptotically
in the $t\to\infty$ limit with power-law (PL) tails, which for non-equilibrium systems
scale as 
\begin{eqnarray} \label{eq:aging}
    \mathcal{R}(t/s) \propto s^{-1-\mathrm{\textit{a}}} (t/s)^{-\lambda_{\mathcal{R}}/Z} \\
    A (t/s) \propto s^{-b}(t/s)^{-\lambda_A/Z} \ ,
\end{eqnarray}
where $t \gg s$, $Z$ is the dynamical, $\lambda_{\mathcal{R}}$ is the auto-response, $\lambda_A$ 
is the auto-correlation exponents and $b$, \textit{a} are the so called aging exponents~\cite{henkel2011non}.
Here $\lambda_A$ and $\lambda_{\mathcal{R}}$ describe the asymptotically large time PL decay
behaviors, while exponents $a$ and $b$ the start time ($s$) scaling dependence of the two-point
functions, happening in non-equilibrium critical systems.

The FD theory states that, if a system is in equilibrium, the response function is related to the 
auto-correlation function of the unperturbed observable by
\begin{equation}
 \mathcal{R}(t,s) = \frac{1}{T} \frac{\partial A(t,s)}{\partial s},
\end{equation}
where $s < t$, and $T$ is the temperature of the thermal bath. 
However, if the system is out of equilibrium a way to measure the difference from equilibrium, 
is through the FD ratio
\begin{equation}\label{eq:FDR}
 \mathcal{R}(t,s) = \frac{1} {T} X(t,s) \frac{\partial A(t,s)}{\partial s} ,\
\end{equation}
where $\frac {1} {T} X(t,s) := \mathrm{FDR}$ characterizes the distance from equilibrium. 
If $\lim_{(t/s)\to\infty} X(t,s)=1$, the system is in equilibrium with temperature $T$.
Conversely, a non-equilibrium system can be characterized by the deviation of 
FDR from a constant value. Note, that this FDT formulation is applicable for stochastic
systems, where at least an annealed noise is present, which can be related to thermalization
in equilibrium. However, in our modeling we neglect the annealed noise and rely on
the chaotic fluctuations, emerging from the nonlinearity. Thus, we can't interpret a constant
ratio in the steady state as a real temperature. This approximation has already 
been addressed in our previous works~\cite{KKIdeco,sniccikk}, where we found that 
weak Gaussian noise does not modify the synchronization transition behavior until it 
remains negligible compered the chaotic "noise" emerging to peak at the transition.
Away from the synchronization transition point this assumption may break down.

In sections (\enquote{\nameref{sec:AC}}), (\enquote{\nameref{sec:AR}}) 
we calculate auto-correlation and response functions and their ratio in 
(\enquote{\nameref{subsec:vFD}}) in case of different modifications of the 
largest currently known connectome, the structural network of 
a fruit-fly for different levels of re-weighting of the edges, corresponding to different brain 
states. These states exhibit different levels of anisotropies.
Before that we provide network analysis of the the structural connectome
and define its states, realizing different levels of edges weights corresponding to
operation of excitatory and inhibitory mechanisms (\enquote{\nameref{sec:scen}}).
Following that we introduce the brain toy model, the Shinomoto-Kuramoto equation (\enquote{\nameref{sec:SK model}})
we used to model the synchronization mechanism of the oscillatory behavior and determine its
critical points for each scenario (\enquote{\nameref{sec:crit}}).

\section{Models and Methods}

\subsection{The topology of the FF connectome} 

Here we present the graph theoretical methods we used to analyze the FF connectome.
Note that these are complementary to those of Ref.~\cite{Lin2024-hz}, but important
for our critical dynamical simulations.

\subsubsection{Community calculations}

The connectome is defined as the structural network of neural connections in 
the brain~\cite{sporns_human_2005}.
For the "original" HB we used the data-set (v1.0.1) from~\cite{down-hemibrain1.0.1}.
The full-fly connectome (FF), version v630, was downloaded from~\cite{down-fullbr}.
These graphs are based on the images of an electron microscopy of the brain of 
an adult \textit{Drosophila melanogaster} fruit-fly~\cite{ZHENG2018730}.
The network statistics of the whole brain were published in \cite{Lin2024-hz}.
That paper is more neuroscience-like, while our approach uses more network 
theory and statistical physics.
Here we show a complementary network analysis, by providing functional fittings,
numerical exponents, dimensions, hierarchical measures, supporting our results 
about our dynamical simulation results.

The adjacency matrix  ($A_{ij}$) is visualized in ~\cite{Flycikk}, where one can 
see a rather homogeneous, almost structureless network however, it is not 
random~\cite{Graphanal}. The degree distribution is much wider than that of random 
Erd\H os-R\'enyi (ER) graphs and exhibits a fat tail. 
The analysis in~\cite{Flycikk} found a global weight distribution 
$p(w)$, with heavy tails. Assuming a PL form, an exponent $-2.9(2)$ could be fitted for the $w > 100$ region.

The modularity quotient of a network with $N$ nodes and average degree $\langle k\rangle$  is defined by \cite{Newman2006-bw}:
\begin{equation}
Q=\frac{1}{N\av{k}}\sum\limits_{ij}\left(A_{ij}-\Gamma
\frac{k_i k_j}{N\av{k}}\right)\delta(g_i,g_j) \ ,
\end{equation}
where  $k_i$, $k_j$ are the node degrees of $i$ and $j$ and the Kronecker delta function $\delta(g_i,g_j)$ is $1$. when nodes $i$ and $j$  are in the same community or $0$ otherwise. 
It depends on the community division of the graph and the maximum value of $Q$ characterizes how modular a network is.

\subsubsection{Graph dimension}

The graph dimension ($d$) is a relevant factor, as it determines if the fluctuations
are relevant near criticality, i.e. the relation of $d$ to the upper critical value $d_c=5$  
of the model. We estimated the effective topological graph dimension $d$, as a generalization
of the Euclidean dimension to graphs, by measuring shortest path chemical distances between nodes 
$r$ via the Breadth-first search ((BFS)) algorithm~\cite{BFSalg}. 
One can give an estimate for the dimension using the definition $N(r) \sim r^d$, 
where we counted the number of nodes $N(r)$ with chemical distance (number of hops) $r$, 
or less from from all nodes of the graphs, which are the seeds of the (BFS) 
and calculated averages over the trials. 

\subsubsection{Spectral dimension}

Graph spectral properties of complex networks have been shown to be particularly relevant to describe 
network structure~\cite{chung1997}.
In case of complex networks it turned out that the so-called spectral dimension ($d_s$) 
provides a better measure for the relevance of fluctuations in the synchronization 
properties~\cite{millan2018,millan2019}. 
This is obtained from the eigenvalue spectrum of the graph Laplacian matrix and can 
be finite in case of complex networks, with infinite topological 
dimensions~\cite{PhysRevLett.134.057401}.
In nutshell, the relevance of the spectral dimension for KM lies on the fact that the linearized KM
describes random walks, described by the graph Laplacian as a transfer matrix of walker hops. 
Following Refs.~\cite{millan2018,millan2019}, we calculate the normalized Laplacian $L$ with elements
\begin{equation}
    L_{ij}=\delta_{ij}-A_{ij}/k_i
    \label{eqs:LnoW}
\end{equation}
for unweighted networks, where $k_i$ denotes the degree of node $i$. For weighted networks, the elements of the normalized Laplacian are given by
\begin{equation}
     L_{ij}=\delta_{ij}-W_{ij}/k_i'\,,
    \label{eqs:LWei}
\end{equation}
where $k_i'=\sum_j W_{ji}$ denotes the weighted in-degree of node $i$. The normalized Laplacian has real eigenvalues $0=\lambda_1 \le \lambda_2 \le \dots \le \lambda_N$, the density of which scales as \cite{burioni1996,millan2018}
\begin{equation}
    \rho(\lambda)\simeq \lambda^{d_s/2-1}
    \label{eqs:dsl}
\end{equation}
for $\lambda\ll 1$, where $d_s$ is the spectral dimension. The cumulative density is then given by
\begin{equation}
    \rho_c(\lambda) = \int_0^\lambda d\lambda'\rho(\lambda') = \frac{2}{d_s} \lambda^{d_s/2}.
    \label{eqs:dsc}
\end{equation}

\subsubsection{Hierarchical graph anisotropy analysis}

Additionally, we have also analyzed the graph anisotropy of communities of the FF using an algorithm proposed by 
Cz\'egel \& Palla ~\cite{Czegel2015}, which measures the level of hierarchy via random walks.
This random walk hierarchy measure is defined as the inhomogeneity of the stationary distribution of random walkers:
\begin{equation}\label{eq:H}
	H = \frac{\sigma(\mathbf{p}^{stat})}{\mu(\mathbf{p}^{stat})} \ ,
\end{equation}
where $\sigma$ is the standard deviation and $\mu$ is the mean of the stationary distribution $\mathbf{p}^{stat}$, thus this is a size-independent quantity.

The stationary distribution of decaying, weighted random walkers is given by:
\begin{equation}
	\mathbf{p}^{stat} = \frac{e^{1/\lambda} - 1}{N} \sum_{n=1}^{\infty} \left( e^{-1/\lambda} \mathbf{T} \right)^n \mathbf{1} \ ,
\end{equation}
where the transition probability matrix is:
\begin{equation}
	T_{ij} = \frac{w_{ij}}{\sum\limits_l w_{lj}} \cdot \frac{w_{ij}}{\sum\limits_l w_{il}} \ ,
\end{equation}
and $\lambda$ is a free parameter.
The update rules for the distribution of $f$ random walkers are:
\begin{align}
	p_i(t) &\rightarrow p_i(t) + \frac{f}{N} \ , \\ 
	p_i(t+1) &= \sum_{j=1}^N T_{ij} p_j(t) \ , \\ 
	p_i(t+1) &\rightarrow \frac{p_i(t+1)}{1+f} \ .
\end{align}

The parameter $\lambda$ controls the characteristic distance a random walker can travel before its weight decays significantly. Smaller $\lambda$ values emphasize local structure, while larger $\lambda$ values allow walkers to explore longer paths. To avoid divergence of the measure, the authors typically set $\lambda = 4$.

The hierarchy measure $H$ captures how unevenly the random walkers are distributed in the stationary state. A large $H$ indicates that the walkers accumulate on a small set of nodes, revealing a strong hierarchical structure.
For chains or stars, $H \to 0$ as $N \to \infty$: walkers spread evenly, indicating no deeply embedded hierarchical levels. For regular trees, $H$ converges to a non-zero constant in the limit of large $N$ , highlighting their deep, branching pyramid structure.

%%%%%%%%%%%%%%%%%%%%%%%%%%%%%%%%%%%%%%%%%%%%%%%%%%%%%%%%%%%%%%%%%%%%%%%%
\subsubsection{Asymmetry in the interactions (reciprocity)} \label{sec:scen}
%%%%%%%%%%%%%%%%%%%%%%%%%%%%%%%%%%%%%%%%%%%%%%%%%%%%%%%%%%%%%%%%%%%%%%%%

As the FF original ({\bf o}) structural connectome links are directed i.e. dendrites or axons, 
we could manipulate the level of asymmetry in the interactions without altering
the topology by changing the egde weights in the following ways: 
\begin{itemize}
    \item ({\bf i}) modeling inhibitions, by flipping signs of weights of $20\%$ of randomly selected edges: $W'_{ij} = -W_{ij}$ .
    \item ({\bf w}) modeling a local homeostasis, by renormalization of incoming link weights : $W'_{ij} = W_{ij}/\sum_j W_{ij}$.
\end{itemize}
In reality both mechanism is related to the synaptic depression or inhibitions. 
We assumed a fixed random distribution of links, but in we know that even when 
inhibitory cell bodies are settled, individual inhibitory synapses constantly form, 
move and disappear over time in response to sensory experience and learning, that
non-adaptive simulations cannot follow precisely.\\ 
Furthermore, part of the nodes FF connectome are glia cells, supporting neurons, 
but following our recent findings on the universal synaptic weight distributions
~\cite{PRRcikk} we assumed that they do not alter our results regarding 
criticality and the violation of the FDR. However, their role should be 
investigated in a deeper data analysis further.

To quantify a measure of link weight asymmetry we introduced the following normalized quantity:
\begin{equation}\label{eq:alpdef}
    \alpha =  \frac{\sum_{i,j}\vert W_{ij}-W_{ji}\vert}{\sum_{i,j}(\vert W_{ij} + W_{ji}\vert)} \ ,
\end{equation} 
where the summation runs over edges, connecting existing links. This form takes into account, 
that in the case of ({\bf i}) negative contributions may cancel out and it also normalizes,
with the sum of absolute value of the link weights of the graph.

%%%%%%%%%%%%%%%%%%%%%%%%%%%%%%%%%%%%%%%%%%%%%%%%%%%%%%%%%%%%%%%%%%%%%%%%
\subsubsection{The Shinomoto-Kuramoto model with external force}\label{sec:SK model}
%%%%%%%%%%%%%%%%%%%%%%%%%%%%%%%%%%%%%%%%%%%%%%%%%%%%%%%%%%%%%%%%%%%%%%%%

We used an extended version of the Kuramoto model (KM) of interacting oscillators~\cite{kura} 
to study the synchronization with a periodic drive, modeling a working-state 
brain~\cite{buendia2021broad,sniccikk}. 
KM is a mathematical framework to describe synchronization of coupled oscillators. 
It is highly relevant for neuroscience as a foundational tool to understand synchronization, 
rhythmic activity, and large-scale brain dynamics. It serves as a simplified, 
mathematically tractable model that bridges microscopic neural behavior with 
macroscopic brain imaging data (EEG, MEG, fMRI)~\cite{10.3389/fnhum.2010.00190}.
Kuramoto showed that an ensemble of phase oscillators interacting through an appropriate functional 
form approximates the long-term behavior of any ensemble of interacting oscillatory systems 
as long as the coupling is weak and the subsystems are nearly identical. This phase reduction 
approach has now become a standard technique in computational neurosciences 
(see, \cite{Ermentrout-Kopell,guckenheimer2013nonlinear,Brown2004}).\\
Computational studies often proceed from the premise that cortical dynamics operate 
in a linearly stable domain, where fluctuations dissipate quickly and show only short memory. 
Studies of human EEG, however, have shown significant autocorrelation at time lags on the scale of 
minutes, indicating the need to consider regimes where non-linearities influence the dynamics. 
Computational models of criticality research, have used simple toy models of statistical 
physics, as well as neurophysiology-inspired population ones.

Currently, we lack models that would accurately describe realistic oscillation-based
functional connectivity and critical dynamics of the brain.
The KM is the simplest model of synchronization dynamics that captures 
the emergence of self-organized synchrony in a system of coupled oscillators.
Furthermore, recently KM could be derived from more realistic, nearly identical quadratic 
integrate-and-fire neurons with both electrical and chemical couplings~\cite{10.1063/5.0075285}.

In KM oscillators with phases $\theta_j(t)$, $j=1,2\dots,N$ are placed on $N$ nodes of a network. 
They evolve according to the following set of dynamical equations
\begin{eqnarray}\label{eq:SK}
\dot\theta_j(t) &=& \omega_j^0+K\sum_{k=1}^{k=N} W_{jk}\sin[\theta_k(t)-\theta_j(t)] \\
\nonumber
&+& F\sin(\theta_j(t)) \ ,
\label{diffeq}
\end{eqnarray}
where  $\omega_j^0$ is the so-called self-frequency of the $j$-th oscillator, 
which is drawn from a Gaussian distribution, with zero mean and unit variance. 
The summation is performed over adjacent nodes, coupled by the $W_{jk}$ weighted 
adjacency matrix. 

In the Shinomoto extension~\cite{10.1143/PTP.75.1105} of the Kuramoto model, 
(SK) we have a periodic force term, proportional to a coupling $F$ to describe external excitation.
This is meant to model the task phase of the brain, in contrast to the KM, which
corresponds to the resting state.
The global coupling $K$ is a control parameter of the model by which we can tune the system between 
asynchronous and synchronous states.
Note, that we omitted stochastic noises, as we found earlier that when they small,
their effect can be substituted by the chaotic noise near criticality, 
coming from non-linearity~\cite{KKIdeco,sniccikk} and make the numerical solution very unstable.
A similar fluctuation-dissipation relation connecting the properties
of fluctuations (correlations) to the stability properties of replica-reliability 
has also very recently studied in~\cite{p5xz-rfv6}.

To locate the synchronization transition we solved the set of
Eqs.\ref{eq:SK} using the Bulirsch-Stoer (BS) adaptive stepper, with step 
size $\delta=0.01$. 
The adaptive BS stepper~\cite{BS1966,BS}, which adjusts step size and degreee of 
function approximation to ensure a local absolute error and relative errors 
$\nu\leq {10^{-9}}$. We used the implementation in 
\texttt{boost::odeint}~\cite{boostOdeInt} with the VexCL backend~\cite{vexcl} 
for support for CUDA GPU graphic cards. 
To locate the synchronization transition we started the system with uniformly distributed random initial 
values of $\theta_j(0)$ phases, but the two-point function calculations were also initialized by $\theta_j(0) \simeq 0$. 
In principle both initial conditions lead to the same steady
state, but finding critical PL decays of $R(t)$ at $K_c$ is hard~\cite{Kurcikk}, because 
$\lim_{t\to\infty} R(t) = 1/\sqrt{N} > 0$. 
In principle this should not be a problem in case of measuring connected two-point functions.
The $t=0$ frequencies were set to be equal to the intrinsic values: 
$\dot \theta_j(0) = \omega_j^0$. 
The Kuramoto phase order parameter 
\begin{equation}\label{ordp}
z(t_k) = r(t_k) \exp\left[i \theta(t_k)\right] = 1 / N \sum_j \exp\left[i \theta_j(t_k)\right] \ .
\end{equation}
was calculated at discrete time steps:
\begin{equation}\label{eq:lnsamp}
t_k = 1 + 1.08^{k}, \ \ \ k=1,2,3...
\end{equation}
to save memory space when we are looking for asymptotic behavior.
Sample averages over different initial $\omega_j^0$ configurations were taken from hundreds of independent realizations
\begin{equation}\label{KOP}
R(t_k) = \langle r(t_k)\rangle
\end{equation}
However, for calculating the two-point functions we had a CPU code available only,
where we used the standard Runge-Kutta (RK4) solver from Numerical Recipes~\cite{NumR}.

In the steady state, which was determined by visual inspection of the $R(t_k,K)$ 
growth results from zero,
we measured the standard deviations: $\sigma(R(t_k\to\infty,K))$ to locate $K_c$
as the fluctuations known to exhibit a maximum there in case of the Kuramoto model.
Alternatively, half values of $R(t_k\to\infty,K))$ also provide an estimate for
$K_c$.

%%%%%%%%%1%%%%%%%%%%%%%%%%%%%%%%%%%%%%%%%%%%%%%%%%%%%%%%%%%%%%%%%%%%%%%%%
\subsubsection{Synchronization transition point determination}
%%%%%%%%%%%%%%%%%%%%%%%%%%%%%%%%%%%%%%%%%%%%%%%%%%%%%%%%%%%%%%%%%%%%%%%%

We determined the synchronization transition points of each variants: (o), (w), (i) 
by early time dependent runs as well as by steady state analysis. 
The time dependent solutions were started from fully phase disordered states 
and followed up to $t_{max} = 300$ time steps.
The sets of equations (\ref{eq:SK}) were solved numerically for $10^3 - 10^4$ independent initial 
conditions and the Kuramoto order parameter is calculated.
The $R(t_k,K)$ is non-zero above the critical coupling strength $K > K_c$, tends to $\propto\sqrt{1/N}$
for $K < K_c$, or exhibits an algebraic growth law at $K_c$:
\begin{equation}\label{escal}
    R(t,N) = N^{-1/2} t^{\eta} f(t / N^{Z}) \ ,
\end{equation}
where $\eta$ is an initial growth exponent in statistical physics of
non-equilibrium critical phenomena~\cite{odorbook}.
Applying a standard local slope analysis~\cite{odorbook}, defined by the logarithmic derivatives of
the growth Eq.~(\ref{escal}) at the discrete time-steps $t_k$, near the transition point
\begin{equation}
    \eta_\mathrm{eff} = \frac {\ln R(t_{k+4}) - \ln R(t_{k})} 
{\ln(t_{k+4}) - \ln(t_{k})} \ ,
\end{equation}
we estimated $K_c$ as well as the exponent $\eta$. Here we used the discrete time derivative 
step size $4$, to lessen fluctuations but we tried with other values (2,3,8) too. 
Here one has to make an compromise between fluctuations and information loss by choosing too large difference steps. T
he value 4 seemed to be an optimal choice. In general the knowledge
of the corrections to scaling allows to select a proper rescaling of the 
horizontal axis and one can observe a linear inflexion curve at $K_c$, 
separating the up and down veering ones, which correspond to the super and 
sub-critical phases. Here one can also extrapolate to the real $\eta$ exponent
in the $t\to\infty$ limit. But in general, we lack the knowledge of the
scaling corrections and we just try the rescaling intuitively, 
assuming $1/t$ usually, which corresponds to the critical behavior
of simplest known non-equilibrium models.

%%%%%%%%%%%%%%%%%%%%%%%%%%%%%%%%%%%%%%%%%%%%%%%%%%%%%%%%%%%%%%%%%%%%%%%%
\subsubsection{Physical aging behavior}
%%%%%%%%%%%%%%%%%%%%%%%%%%%%%%%%%%%%%%%%%%%%%%%%%%%%%%%%%%%%%%%%%%%%%%%%

For analyzing the violation of FDT  we measured the following difference 
function in Eq.(\ref{eq:R}) for the responses: 
\begin{equation}\label{Rdef}
    E_h(t,s) = 1/N
    \sum_{i=1}^N \cos(\theta^{'}_i(t)-\theta_i(t)) 
\end{equation}
of the angle variables $\theta_i(t)$ of the original and the perturbed replicas
$\theta^{'}_i(t)$, as the response for an external small perturbation 
committed at the start time $s$.
Similarly, the real valued auto-correlations of oscillators are calculated as in
\cite{Acebron}:
\begin{equation}\label{Adef}
A(t,s) =  1/N
    \sum_{i=1}^N  \cos (\theta_i(t)-\theta_i(s)]) \  .
\end{equation}

In this study we applied single random perturbations at all sites~\footnote{Single node 
perturbations provided negligible effects in general.}: 
$\theta'_i = \theta_i +  h \xi_i$, where $\xi_i$ is a random variable, 
drawn from a zero centered Gaussian distribution and measured the auto-responses by
Eq.(\ref{eq:R}) with $h=0.001$ and determined it's start time derivatives.

%%%%%%%%%%%%%%%%%%%%%%%%%%%%%%%%%%%%%%%%%%%%%%%%%%%%%%%%%%%%%%%%%%%%%%%%
\subsubsection{Measuring two-point functions}
%%%%%%%%%%%%%%%%%%%%%%%%%%%%%%%%%%%%%%%%%%%%%%%%%%%%%%%%%%%%%%%%%%%%%%%%

The auto-correlator (and the auto-response) function calculations  
were performed by a C code, which computes these functions up to $t_{max} = 2000$
with the following start time values $s=0.5$,$1$,$2$,$4$,$10$,$20$,$40$,$50$,$100$,$200$ in parallel. 
Throughout this paper the time is measured by full graph Kuramoto updates of the solver.  
Numerical solutions were re-run for hundreds of independent initial self-frequency distributions, 
usually starting from phase synchronized states and averaged over finally.
We also performed tests by starting from random phase distributions, but in that case the fluctuation 
effects were stronger. Note, that we avoided the addition of stochastic noise as it makes the 
solutions numerically more instable, instead a chaotic "noise" appears 
in the system naturally due to the strong nonlinearities. 
This introduces strong quenched heterogeneity besides the topological ones.

From these extensive calculations we determined the $s$ derivatives of the correlators:
$d A(t,s)/ ds := \dot A(t,s)$. For simplicity we reduced the two-point functions
to be variables of $t/s=y$, which is expected to be the scaling variable in aging system at 
criticality~\cite{henkel2011non}, providing the autocorrelation and auto-response
exponents in the $y\to\infty$ limit, as shown in Eq.~(\ref{eq:aging}). 

\section{Results}

\subsection{Topology of the full fruit-fly connectome (FF)}

In this section we present results related to the graph theoretical
analysis of the FF connectome, for better understanding the dynamical
KM simulation outcomes. As we assume brain criticality we consider those metrics,
which are known to be relevant for non-equilibrium critical systems~\cite{odorbook}.

\subsubsection{Communities}

The $Q$ value, characterizing modularity (see Models and Methods) is not 
independent of the community detection method. 
If our detection method produces lower modularity than the maximum achieved, 
it means our algorithm has fallen behind others. 
Community detection algorithms based on modularity optimization will get 
the closest to the actual modular properties of the network. 
We calculated and compared the modularity quotients, using community structures 
of the FF and the HB~\cite{down-hemibrain1.0.1},
by the Louvain method \cite{Blondel2008}. The results for 1000 iterations on these networks are:
$Q_{HB} \approx 0.613(6)$, $Q_{FF} \approx 0.649(4)$,
similar to each other despite all the differences and very far from the highly
modular ($Q\approx0.91$--$0.92$) human connetomes. 
%%%%%%%%%%%%%%%%%%%%%%%%%%%%%%%%%%%%%%%%%%%%%%%%%%%%%%%%%%%%%%%%%%%%%%%%%
\begin{figure}[h]\centering
\includegraphics[width=0.7\columnwidth]{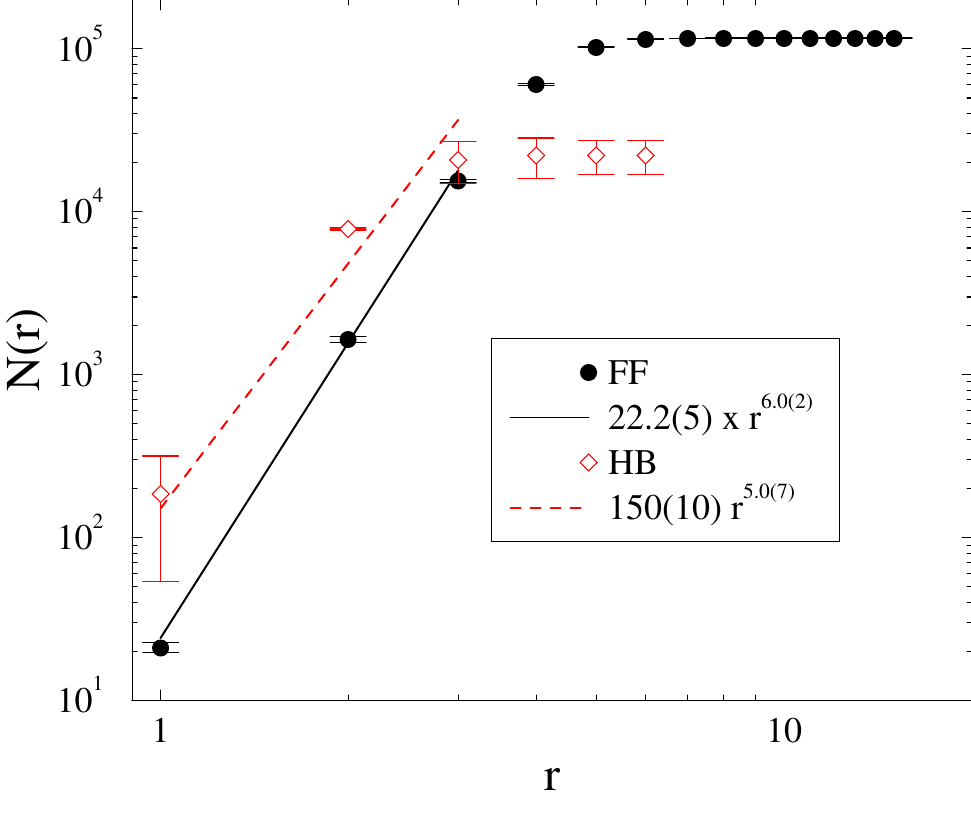}
\caption{Comparison of the topological dimension measurements of the full fly (FF) and the HB. 
Bullets show the number of nodes $N(r)$, within graph (chemical) distances $r$ in case of the FF, 
while hollow diamonds the same in case of the HB. 
Lines display least square PL fits \red{for $r\le 3$}, with exponents corresponding to the 
topological (graph) dimensional estimates. We can observe that this dimension is higher
than 5, where the critical dynamical fluctuations are irrelevant, especially in case of the FF.
But an early finite size cutoff makes these fitting estimates rather uncertain.}
\label{fig:fly-dim}
\end{figure}
%%%%%%%%%%%%%%%%%%%%%%%%%%%%%%%%%%%%%%%%%%%%%%%%%%%%%%%%%%%%%%%%%%%%%%%%%

The FF connectome contains $N_{FF}=124 891$ nodes and $L_{FF}=3 794 615$ edges, 
out of which the largest single connected component has $N_{FF}=115 661$ nodes and $L_{FF}=3 719 871$ links. 
Note that the FF connectome is 5 times bigger, yet the number of links that connect them is not that much larger. 
This results an average node degree of $\langle k\rangle= 64.323$ (for the in-degrees it is: $32.161$), 
the average weighted degree is $\langle w_{FF}\rangle= 557.241$.
Similarly, to the HB for the FF we found PL tailed global weight distributions, 
characterized by the exponent: $-3.0(1)$, even if glia cells were filtered out. 
On the other hand the in/out degree (local, weight) 
distributions could be better fitted via log-normal distributions~\cite{PRRcikk}.

\subsubsection{Graph (topological) dimension}

These measurements are explained in the Section Models and Methods.
Note, that finite size cutoff happens very early. Fig.~\ref{fig:fly-dim} shows the 
comparison of dimensions, obtained for the HB and the FF. 
While we estimate $d\simeq 5.0(7)$ for the HB, for the FF we obtain an even higher dimension: 
$d\simeq 6.0(2)$. 
Note, that due to the small world property the finite size cutoff occurs at small 
chemical distances, thus our $d$ estimates, based on a scaling assumption is rather 
uncertain and provide just a hint for $d \ge 5$ based on these fittings.
These dimensions suggest that dynamical models on the fly connectomes exhibit mean-field 
critical behavior, as the upper-critical dimension is $d_c = 5$ in case of the 
Kuramoto model~\cite{HPCpre,llcikk}. 
However, this $d_c$ has been proven in case of regular, finite dimensional lattices,
so in the followings we consider the spectral dimension, as a more relevant factor in case of
heterogeneous complex networks and KM.

\subsubsection{Spectral dimension}

\begin{figure}[h]\centering
\includegraphics[width=0.7\columnwidth]{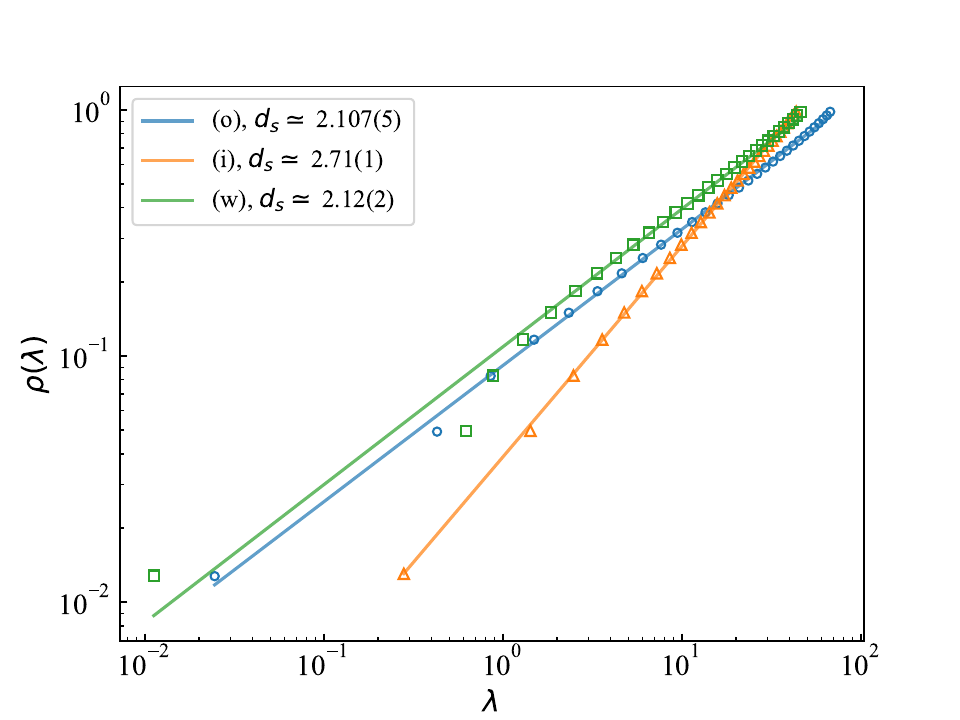}
\caption{A characterization of the synchronization dynamics of the networks based on their spectral dimensions $d_s$, obtained from the scaling behavior of the lowest eigenvalues ($\lambda$) of the normalized Laplacian matrix. The figure displays the eigenvalue density $\rho(\lambda)$ for the $2500$ lowest eigenvalues of the FF connectome across the three scenarios: \textbf{o} (circles), \textbf{i} (squares), \textbf{w} (triangles). The density $\rho(\lambda)$ is computed using equal quantile binning in logarithmic space across $N = 30$ bins. As we can see, contrary to the graph dimensions $d$, all spectral dimensions satisfy $d_s < d_c = 5$, suggesting non-mean-field type of synchronization transitions.}
\label{fig:FullFly-spect}
\end{figure}

As discussed in Sect. Models and Methods we calculated the eigenvalues of the Laplacian $L_{ij}$ of the FF connectome graph.
As scaling behavior of the density of eigenvalues, i.e. Eqs.~\eqref{eqs:dsl} and \eqref{eqs:dsc} are expected to hold for small 
$\lambda$-s, we extract the densities from the first 2500 smallest eigenvalues for ease of computation without loss of generality.
With this method the $d_s$ of the HB connectome has already been estimated~\cite{chimera} and $d^o_s=3.37$, 
for the original, $d^w_s=2.34$, for the weighted dimensions were obtained.
The full FF connectome contains $N_{ff}=124.891$ nodes, requiring large computational efforts, 
therefore, we used the thick-restart Lanczos \cite{Kesheng2000} with sparse matrix representation and accelerated 
with graphics processing unit (GPU), using the latest version (13.4.0) of the CuPy extension. 
As we can see on Fig.~\ref{fig:FullFly-spect} for (o) and (i) we get $d^o_s=2.107(5)$, $d^i_s=2.71(1)$ 
respectively, within error margins of the least squares fitting, while for (w) also a larger value 
than $d^o_s$ is obtained, similar to HB case:  $d^w_s=2.12(2)$. 
Important to note that these dimensions are smaller than the $d_c$ threshold of the mean-field 
behavior~\cite{millan2018}, thus we may observe non-mean-field behavior of the KM at the synchronization 
transition. 

\subsubsection{Asymmetry of the FF connectome variants}\label{sec:FFaVAR}

To characterize the global graph anisotropy of the edge weight modified versions of 
the FF we introduced a measure in Sect. Models an Methods, 
which provides the following values of $\alpha$ for the scenarios:
\begin{itemize}
    \item $\alpha = 0.769$, for the original FF (o) 
    \item $\alpha = 0.907$, for case (i), with $20\%$ random negative weights    
    \item $\alpha = 0.663$, for the renormalized weights case (w)
\end{itemize}
We can see that the renormalized connectome (w) is the less asymmetric, followed by 
the original (o), and finally (i) exhibits the largest $\alpha$ value. 
Independently the scenarios about $3.7\%$ of the directed
links have a backward pair, a reciprocity level fixed by the connectome
structure that we did not want to alter.
In this study we shall compare the dynamical behavior of the
Kuramoto model on these networks versions.

\subsubsection{Graph hierarchy}

\begin{figure}[h]\centering
\includegraphics[width=0.7\columnwidth]{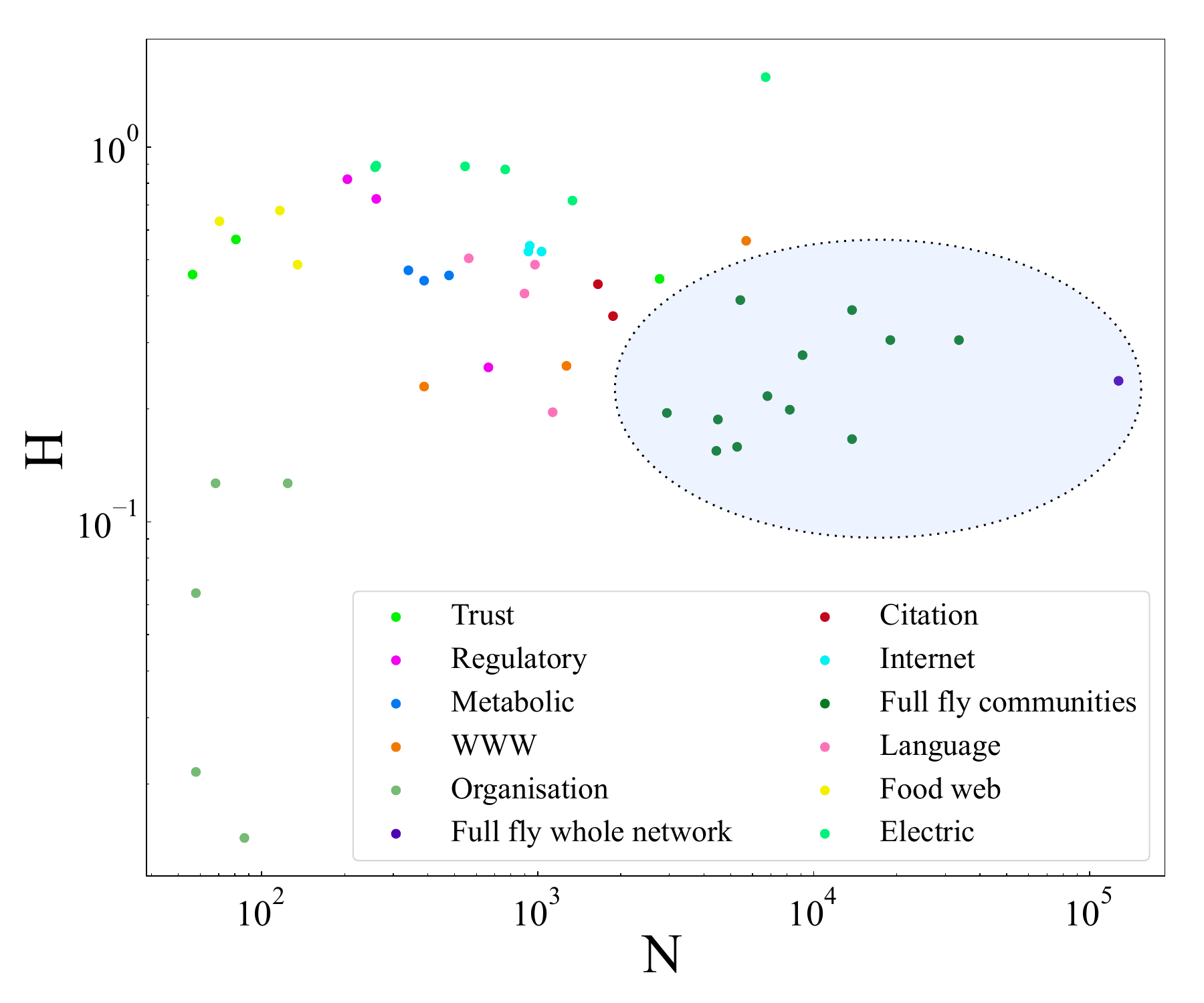}
\caption{Here, we show the random walk hierarchy measure $H$ (Eq. (\ref{eq:H}))
of the FF connectome (rightmost purple point) and all of its communities  
(inside the transparent ellipse), in comparison with other real 
world networks from~\cite{Czegel2015}. 
The modules are noticeably larger than former empirical networks considered 
before and they exhibit rather low level of hierarchy. }
\label{RWHfig}
\end{figure}

We calculated the $H$ value of each community of the FF network.
We can see on Fig.\ref{RWHfig} that the sizes of the FF communities 
are larger than those of the literature reference ones, 
but their hierarchy values $H$ are somewhat smaller. The encircled region
characterizes the hierarchy of the whole FF graph. 
This result also contributes to the debate whether brain networks are hierarchical
modular (HMN) or just simply modular.

%%%%%%%%%%%%%%%%%%%%%%%%%%%%%%%%%%%%%%%%%%%%%%%%%%%%%%%%%%%%%%%%%%%%%%%%%
%%%%%%%%%%%%%%%%%%%%%%%%%%%%%%%%%%%%%%%%%%%%%%%%%%%%%%%%%%%%%%%%%%%%%%%%%

\subsection{Dynamical analysis}

In this section we present results of extensive KM dynamical 
simulations on different versions of the FF graph. First we locate
the synchronization transition point in Sec.~\ref{sec:crit} following that
we perform auto-correlation and auto-response calculations shown in
Sects.~\ref{sec:AC},~\ref{sec:AR}. Finally in Sect.~\ref{subsec:vFD} we
compare these and calculate the FDR. In Sec.~\ref{sec:crit} we also
show results for the Shinomoto-Kuramoto model, describing the effect 
of periodic external drive on the fluctuations in case of the 
scenario (w).

\subsubsection{Synchronization transition point results}\label{sec:crit}

\begin{figure}[h]\centering
\includegraphics[width=0.7\columnwidth]{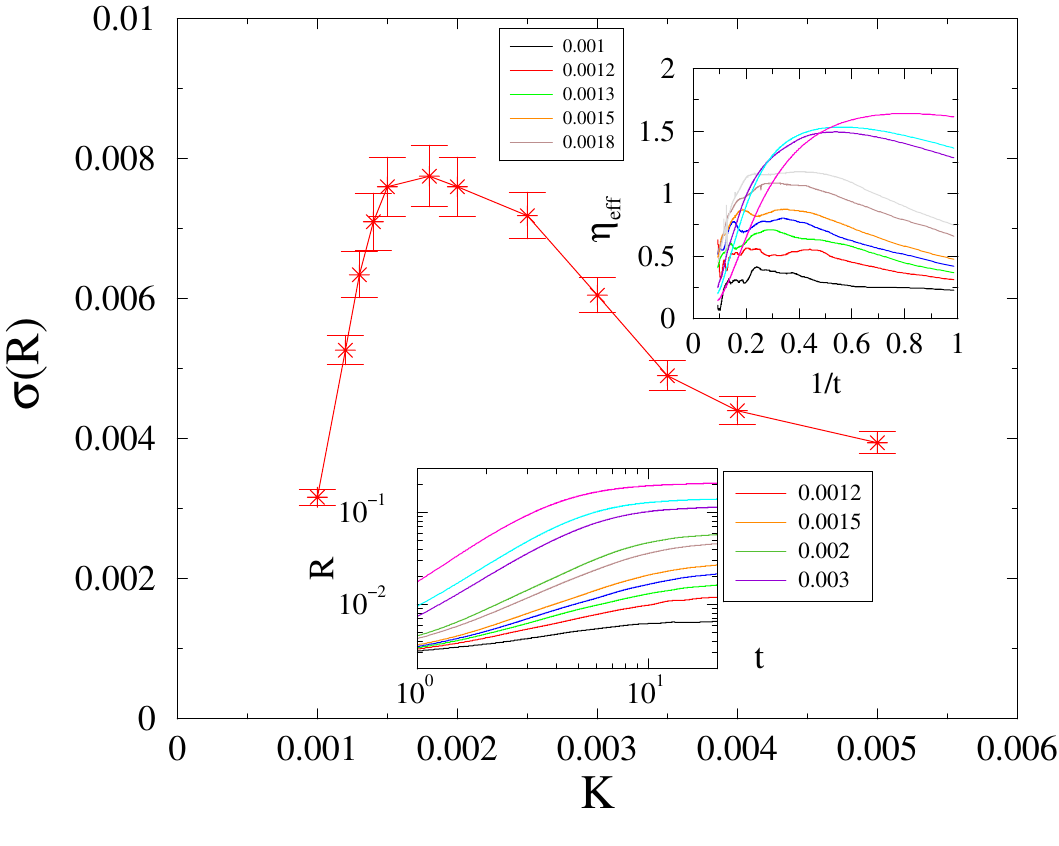}
\caption{Steady state fluctuation (standard deviation) peak of the $R$, as the function of 
the global coupling, for the case {\bf o}.
The lower inset shows the growth of $R(t)$ from incoherent initial states,
for $K=$0.001, 0.0012, 0.0013, 0.0014, 0.0015, 0.0018, 0.002, 0.003, 0.0035, 0.005, 
(bottom to top curves), while the right inset the local slopes of the same
data. Legends correspond to typical values from the above list.
One can locate a transition at $K_c = 0.0015(5)$, where the local slopes tend 
to the $\eta \simeq 0.9(1)$ via a straight line approximation in the $t\to\infty$ limit,
suggesting non-mean-field behavior ($\eta_{MF}=0.75$).
The tiny fluctuations of the lines of time dependency are not shown as 
they would exclude to see their curvatures.}
\label{beta_Flynw}
\end{figure}

In case of scenario ({\bf o}) for the Kuramoto equation ($F=0$), we estimated the 
synchronization transition point to be at $K_c=0.0015(5)$, via finding
an inflexion curve on the $R(t)$ growth plot (bottom inset of  Fig~\ref{beta_Flynw}), 
and by the fluctuation peak of $\sigma(R)$ (Fig~\ref{beta_Flynw}). 
Error margins are estimated by comparing results at subsequent values.
Note, that mean-field scaling is characterized by $\eta_{MF}=0.75$~\cite{cmk2016,llcikk}, 
as compared to the $\eta \simeq 1$ value, one can read-off from the 
right inset  of Fig~\ref{beta_Flynw} at $K_c$, by extrapolating the local slopes
to the $1/t\to 0$ limit. This suggests a deviation from the mean-field 
critical behavior of the KM on the original FF graph, in agreement with the
low spectral dimension.

%%%%%%%%%%%%%%%%%%%%%%%%%%%%%%%%%%%%%%%%%%%%%%%%%%%%%%%%%%%%%%%%%%%%%%%%%
%%%%%%%%%%%%%%%%%%%%%%%%%%%%%%%%%%%%%%%%%%%%%%%%%%%%%%%%%%%%%%%%%%%%%%%%%

\begin{figure}[h]\centering
\includegraphics[width=0.7\columnwidth]{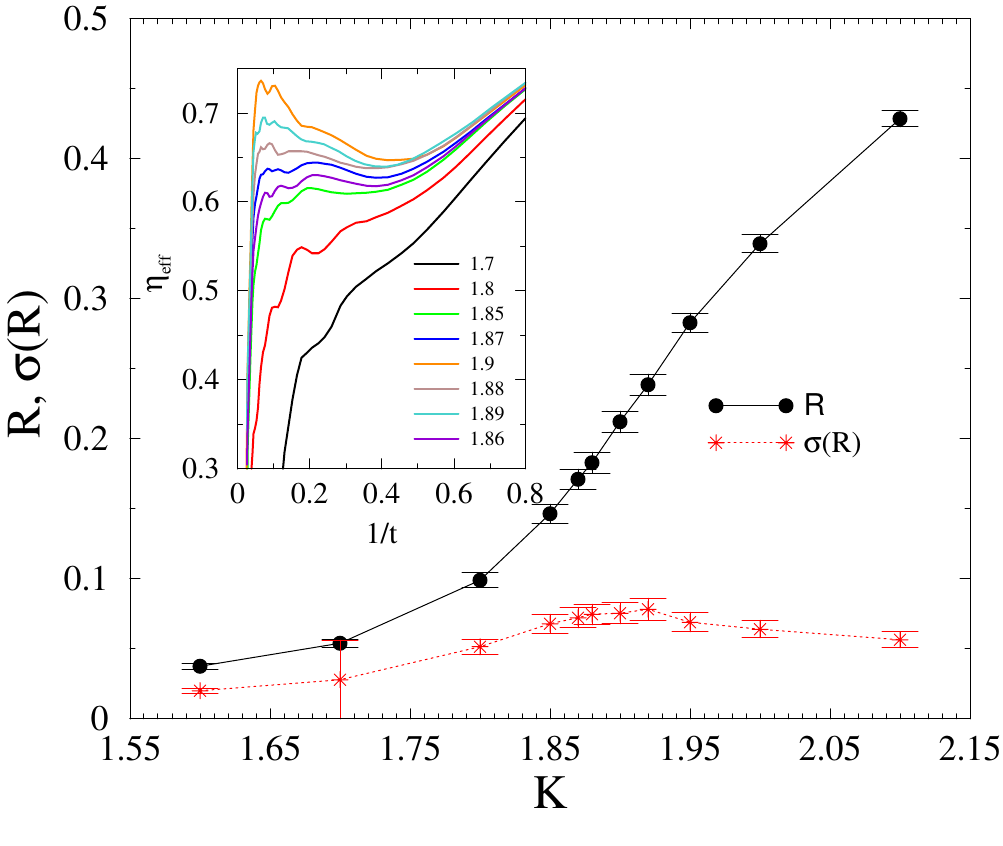}
\caption{The Kuramoto order parameter (black bullets) and its fluctuation peak (standard deviation) 
$\sigma(R)$ (red stars) as the function of the global coupling, for the case {\bf (w)}.
The inset shows the local slopes of the growth of $R(t,K)$, from incoherent initial states, for
$K$=1.7, 1.8. 1.85. 1.86, 1.87, 1.88, 1.89, 1.9. (bottom to to curves). 
These coupling values are also shown by the legends. 
This suggests a transition at $K_c = 1.88(2)$, where the local slopes tend 
to $\eta=0.65(2)$, slightly below the mean-field value $\eta_{MF}=0.75$ 
via a straight line in the $t\to\infty$ limit. This is in agreement with the fluctuation peak 
value at $K_c \simeq 1.9$ within error margin as one can see at the bottom of the figure.
The tiny fluctuations of the lines of time dependency are not shown as they would 
exclude to see their curvatures.}
\label{growthfig}
\end{figure}

In case of scenario ({\bf w}), the transition of the Kuramoto model ($F=0$) happens at a 
much larger coupling: $K_c=1.90(2)$, estimated by the same method, displayed in 
Fig.~\ref{growthfig}.

\begin{figure}[ht]\centering
\includegraphics[width=0.7\columnwidth]{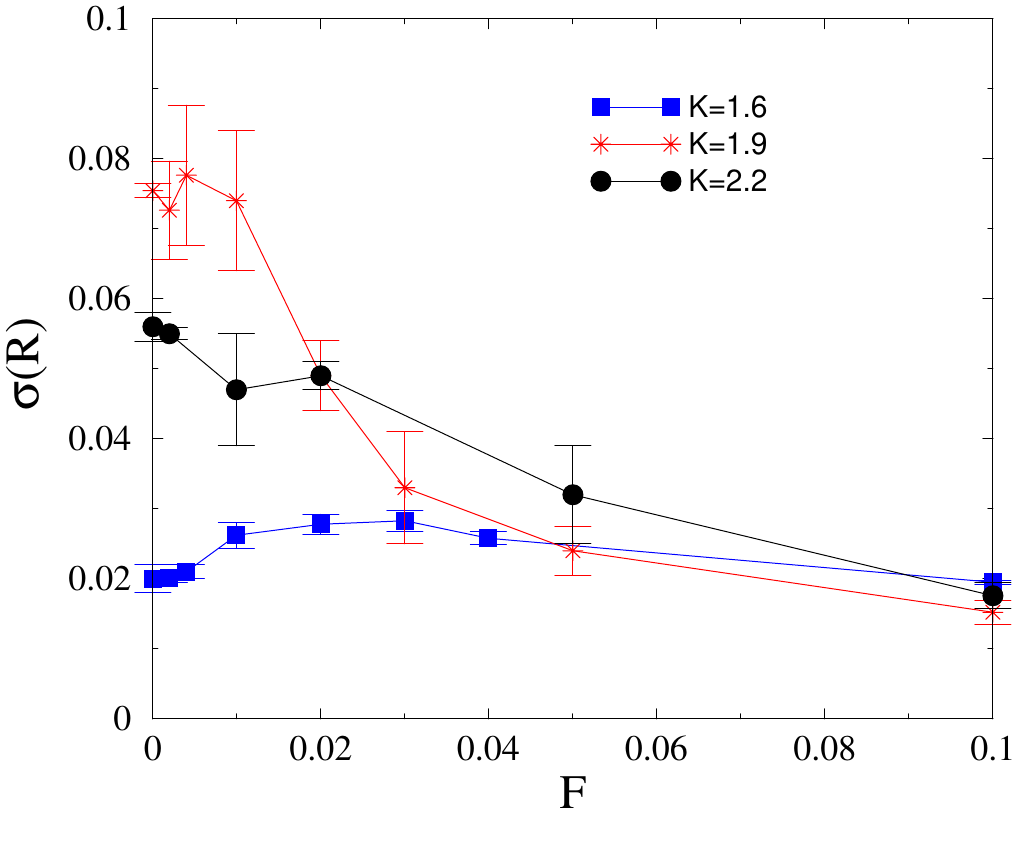}
\caption{Dependence of the steady state SK order parameter $\sigma(R)$
on the global coupling $K$ and on the force $F$ for network {\bf w}). 
One can observe a maximum at $K \simeq 1.9$ and $F \simeq 0$, i.e at 
synchronization transition of the resting state of the brain.} 
\label{betafig}
\end{figure}

We have also investigated the $F$ and $K$ dependence of the $\sigma(R)$, i.e the SK order 
parameter and found that the largest fluctuation value is reached at 
$K_c \simeq 1.9$ close to the critical point without external force, as shown 
on Fig.~\ref{betafig}.
%%%%%%%%%%%%%%%%%%%%%%%%%%%%%%%%%%%%%%%%%%%%%%%%%%%%%%%%%%%%%%%%%%%%%%%%%
%%%%%%%%%%%%%%%%%%%%%%%%%%%%%%%%%%%%%%%%%%%%%%%%%%%%%%%%%%%%%%%%%%%%%%%%%
This means that maximal fluctuations and sensitivity of the system 
happen right at the critical point in the resting state. 
Periodic, external excitations decrease this.

\begin{figure}[h]\centering
\includegraphics[width=0.7\columnwidth]{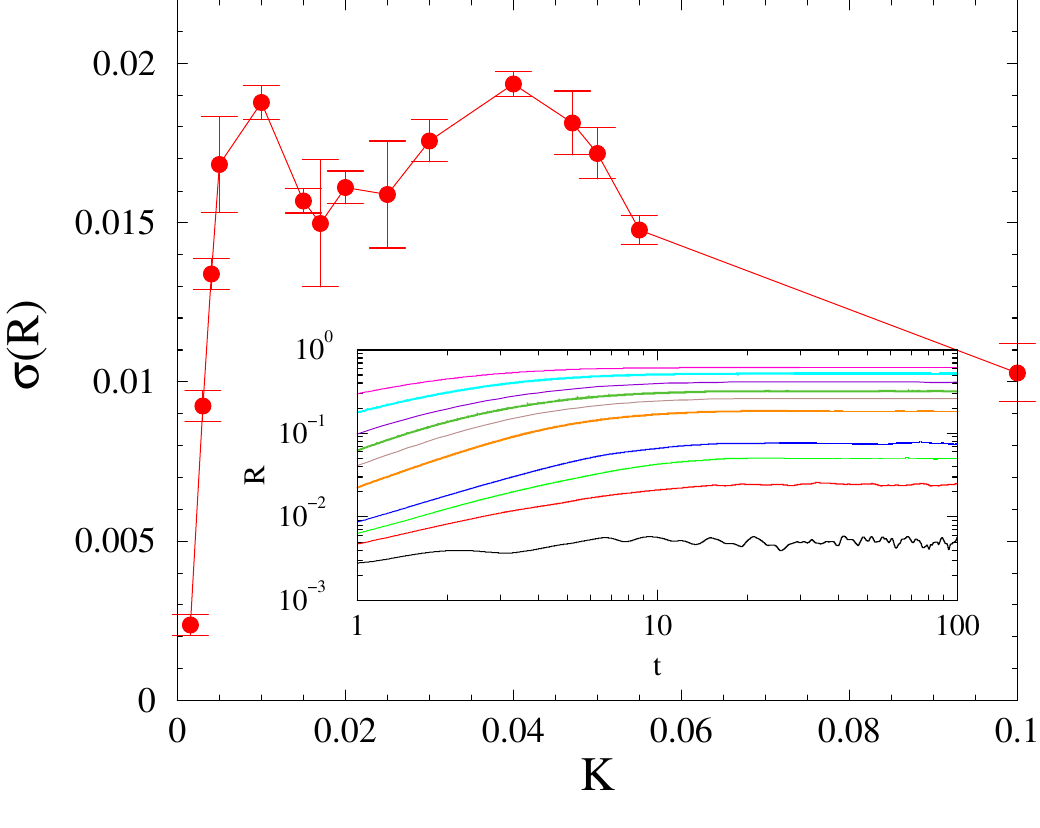}
\caption{Steady state fluctuation peak $\sigma(R(K))$ as the function 
of the global coupling, for the case {\bf (i)}.
The inset shows the growth of $R(t)$ for different global coupling values: 
$K=$ 0.0015, 0.003, 0.004, 0.005, 0.01, 0.015, 0.02, 0.03, 0.05, 0.1 (bottom to top curves). 
One can observe two fluctuation peaks: $K_c = 0.010(1)$ and $K = 0.040(1)$. 
The temporal behavior in the inset suggests dynamical scaling (straight line) near 
the lower peak coupling value (orange/blue lines) at $K \simeq 0.01$. 
The tiny fluctuations of the lines are not shown as they would exclude to see 
their curvatures.}
\label{beta_Flynwi}
\end{figure}

Interestingly, in the case of scenario ({\bf i}), we can observe two fluctuation peaks, 
which should be the consequence of this highly heterogeneous, directed, glassy network. 
By considering the inflection point of the growth we provide an estimate for
the synchronization $K_c\simeq 0.01(5)$ as shown in Fig.\ref{beta_Flynwi}.
That means we need stronger couplings to get synchronization with inhibitory
links, than in case original raw graph.
Understanding of this anomalous behavior would require further numerical studies. 

\subsubsection{Auto-correlation results}\label{sec:AC}
\addcontentsline{toc}{section}{Auto-correlation results}

\begin{figure}[h]
\centering
\includegraphics[width=0.75\columnwidth]{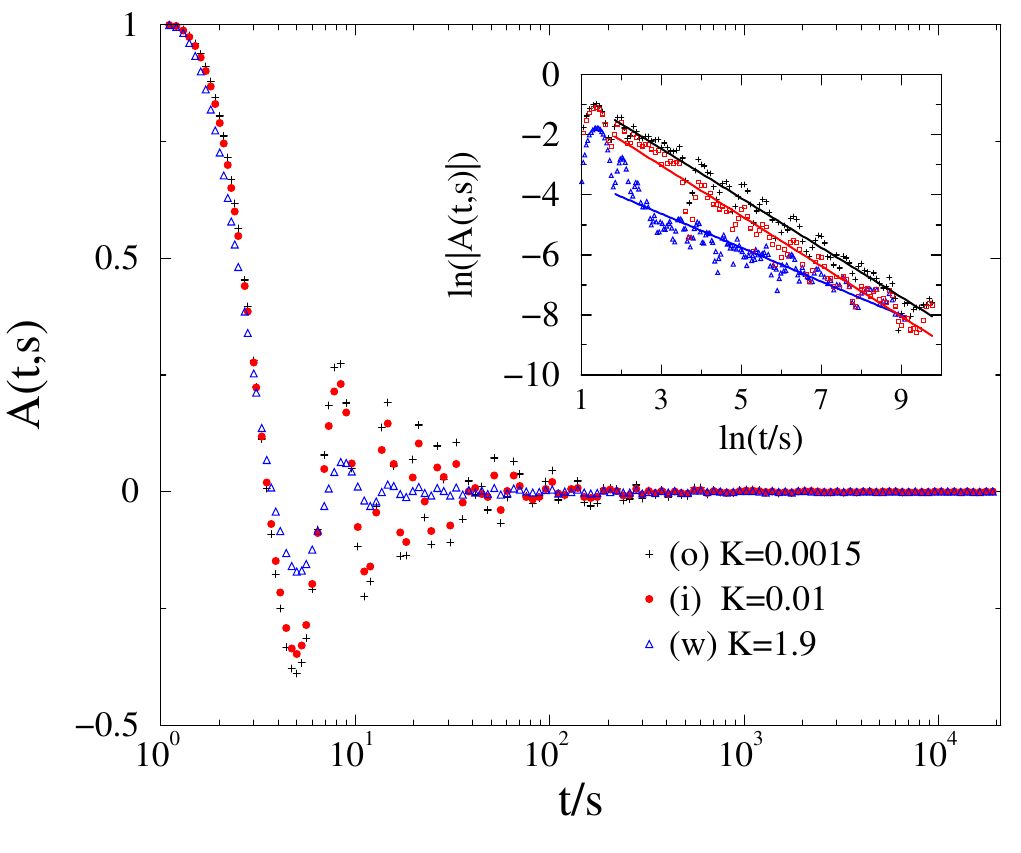}
\caption{Critical auto-correlation functions for $s=1$ in the Kuramoto model, 
for different anisotropy scenarios labeled by the legends in case of 
random phase initial conditions. Inset: Absolute value of the same on logarithmic scales. 
Dashed lines show PL fits for the tails for $t/s$ > 2. 
Independent goodness of fit tests are provided in the Appendix II.}
\label{fig:Aabs}
\end{figure}

The auto-correlation functions were found to exhibit damped oscillatory behavior after averaging 
over thousands of independent samples, corresponding to different self-frequencies 
of nodes and random initial phases, as shown on Fig.\ref{fig:Aabs}
After making the absolute values of $A(t,s=1)$ we can observe an asymptotic PL decay 
of the tails as shown in the inset of Fig.\ref{fig:Aabs}, which is in agreement with 
the critical point expectations for $y := t/s\to \infty$.

The fitted tails for $\ln(y) > 5$ provide the following functions: 
$A_T(t,s=1) = 1.02\times t^{-0.82(2)}$ for (o);
$A_T (t,s=1) = 0.6 \times t^{-0.83(2)}$ for (i), while for the weighted 
case (w) we obtained: $A_T (t,s=1) = 0.05 \times t^{-0.66(5)}$. 
For estimating these numerical values in general in this paper we used least-square 
error fitting and provided standard errors as margins. Note, however that the
fitted exponent values are not expected to agree with the real $\lambda_A/Z$ exponents 
due to application of the absolute value and the Figure serves to visualize the 
complex scaling behavior in this system.

\begin{figure}[ht]
\centering
\includegraphics[width=0.7\columnwidth]{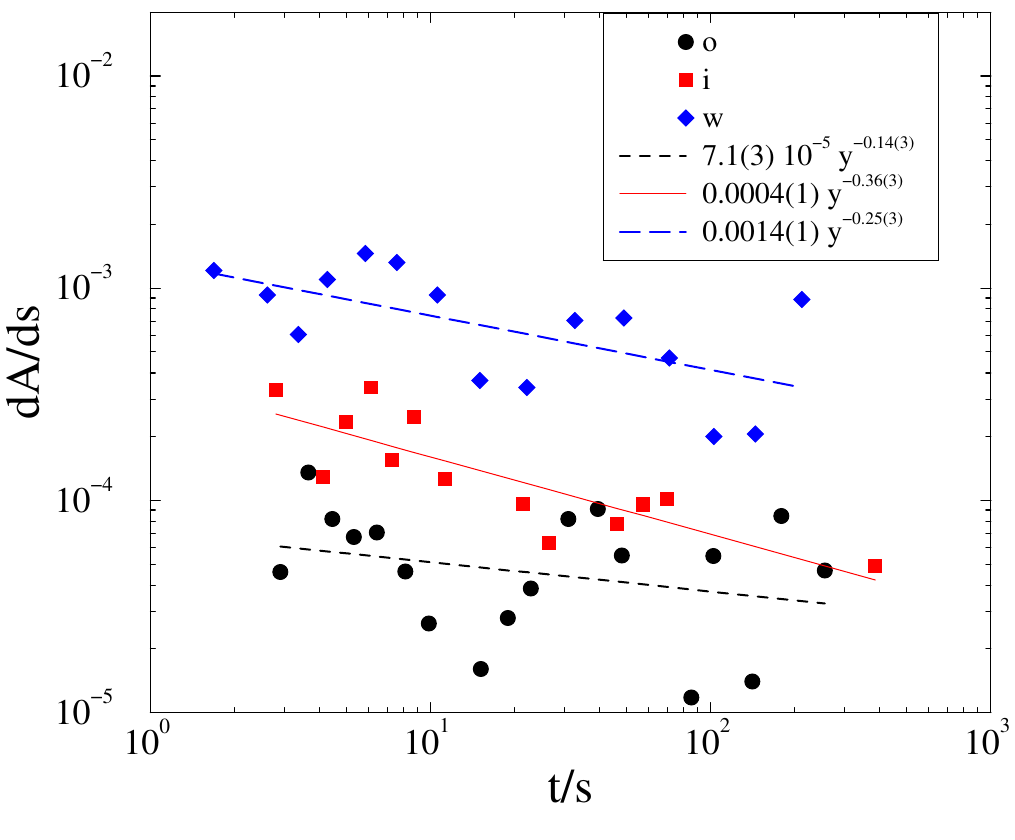}
\caption{Summary of the $s$ derivatives of the critical auto-correlation functions: $\dot A(t,s)$ 
for scenarios ({\bf o}) (bullets), ({\bf i}) (boxes) and ({\bf w}) (diamonds), deduced from the largest time 
results: $t=258$ and $516$ displayed by the legends. Lines show power law fittings 
assuming $A_0\times y^{A_1}$ forms for scenarios: ({\bf (o)}), ({\bf i}), ({\bf w}) 
(bottom to top curves).
Independent goodness of fit tests are provided in the Appendix II.}
\label{fig:dA-Kur-w}
\end{figure}

After calculating the discrete differences in $s$ as described in the Sect.
Models and Methods, we performed rescaling with $s$ and calculated the 
$\dot A(y)$ for each connectome variant, by averaging over the largest 
$t=258, 516$ values to obtain the $y\to\infty$ asymptotic scaling. 
Figure~\ref{fig:dA-Kur-w} shows these results on a log-log plot, following 
application of running averages of sizes of 4-8 points. 
One can see relative oscillations become stronger with $\alpha$ as in case of the 
pure auto-correlations (Fig.\ref{fig:Aabs}).
Assuming log-periodic oscillations we fitted PL-s of the form: 
$\dot A_T(y) = A_0\times y^{A_1}$, by least squares errors.
We used these oscillation filtered  functions to calculate the FDR-s later in 
Sect.~\ref{subsec:vFD}.

\subsubsection{Auto-response results}\label{sec:AR}

We found different decays for different scenarios. 
This is in agreement with the spectral dimension finding
that these connectomes exhibit $d_s < 4$, thus the Kuramoto 
model on these connectomes does not show mean-field like scaling behavior.
\begin{figure}[h]\centering
\includegraphics[width=0.7\columnwidth]{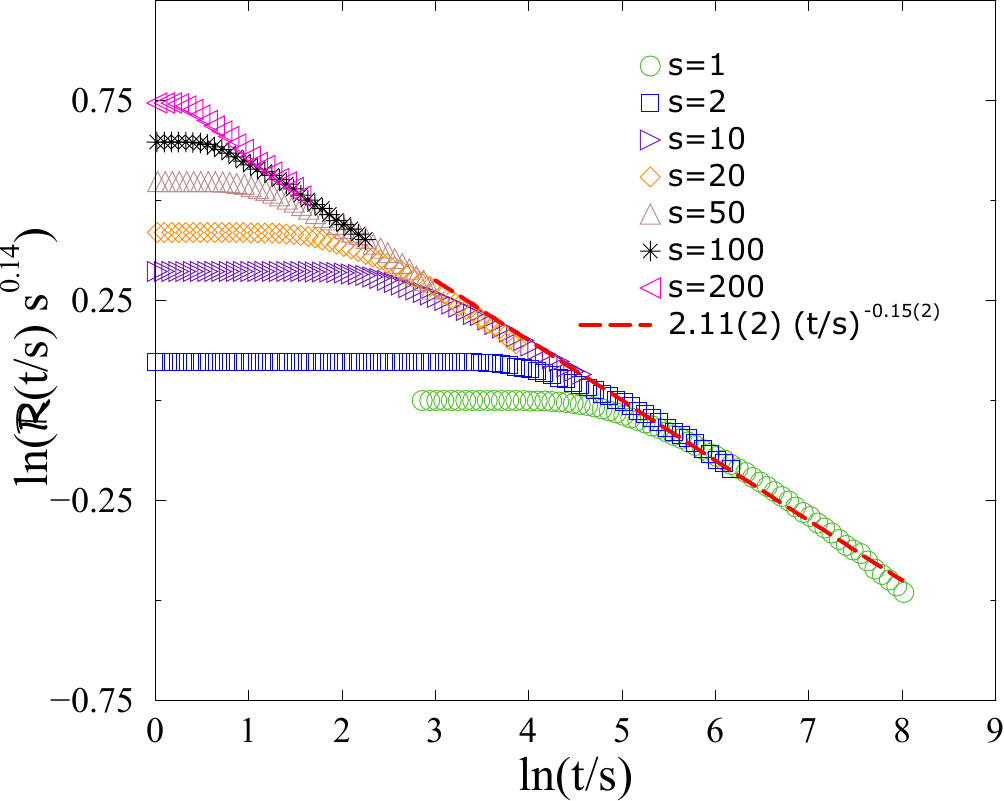}
\caption{Scaling collapse of the auto-responses  ${\cal R}(t,s)_o$ in the Kuramoto model 
for different $s$ values indicated by the legends, in case of graph scenario 
({\bf o}) at the estimated critical point $K_c=0.0015$. 
The dashed line shows a PL fitting for the tails: $\ln(t/s) > 3$. 
Independent goodness of fit tests are provided in the Appendix II.}
\label{fig:R-Kur-o}
\end{figure}

\begin{figure}[h]\centering
\includegraphics[width=0.7\columnwidth]{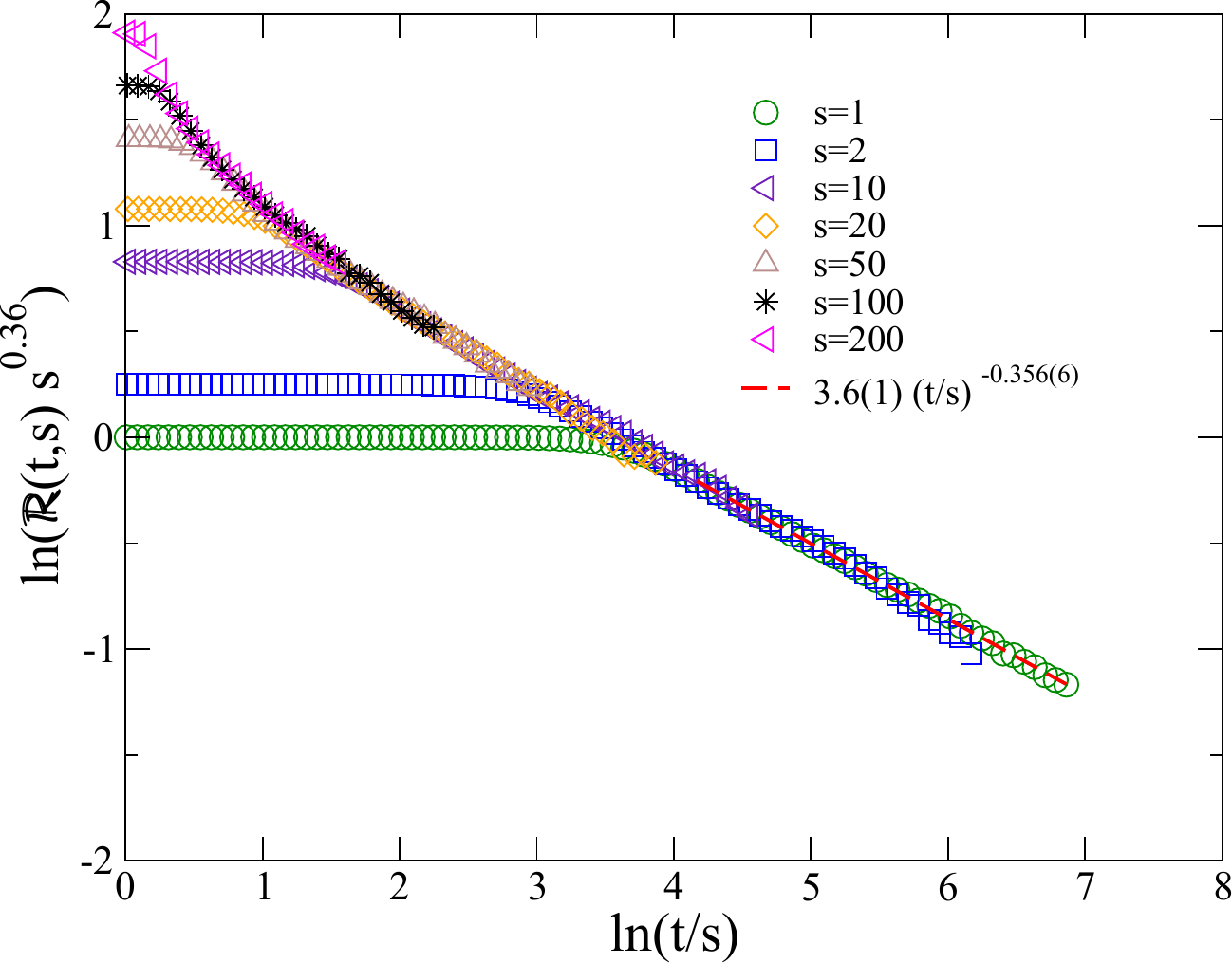}
\caption{Scaling collapse of the auto-responses ${\cal R}(t,s)_i$ in the Kuramoto model 
for different $s$ values indicated by the legends, in case of scenario ({\bf i}) 
	at the critical point: $K_c=0.01$. 
The dashed line shows PL fitting for the tail: $\ln(t/s) > 4.$ 
Independent goodness of fit tests are provided in the Appendix II.}
\label{fig:R-Kur-i}
\end{figure}

\begin{figure}[h]\centering
\includegraphics[width=0.7\columnwidth]{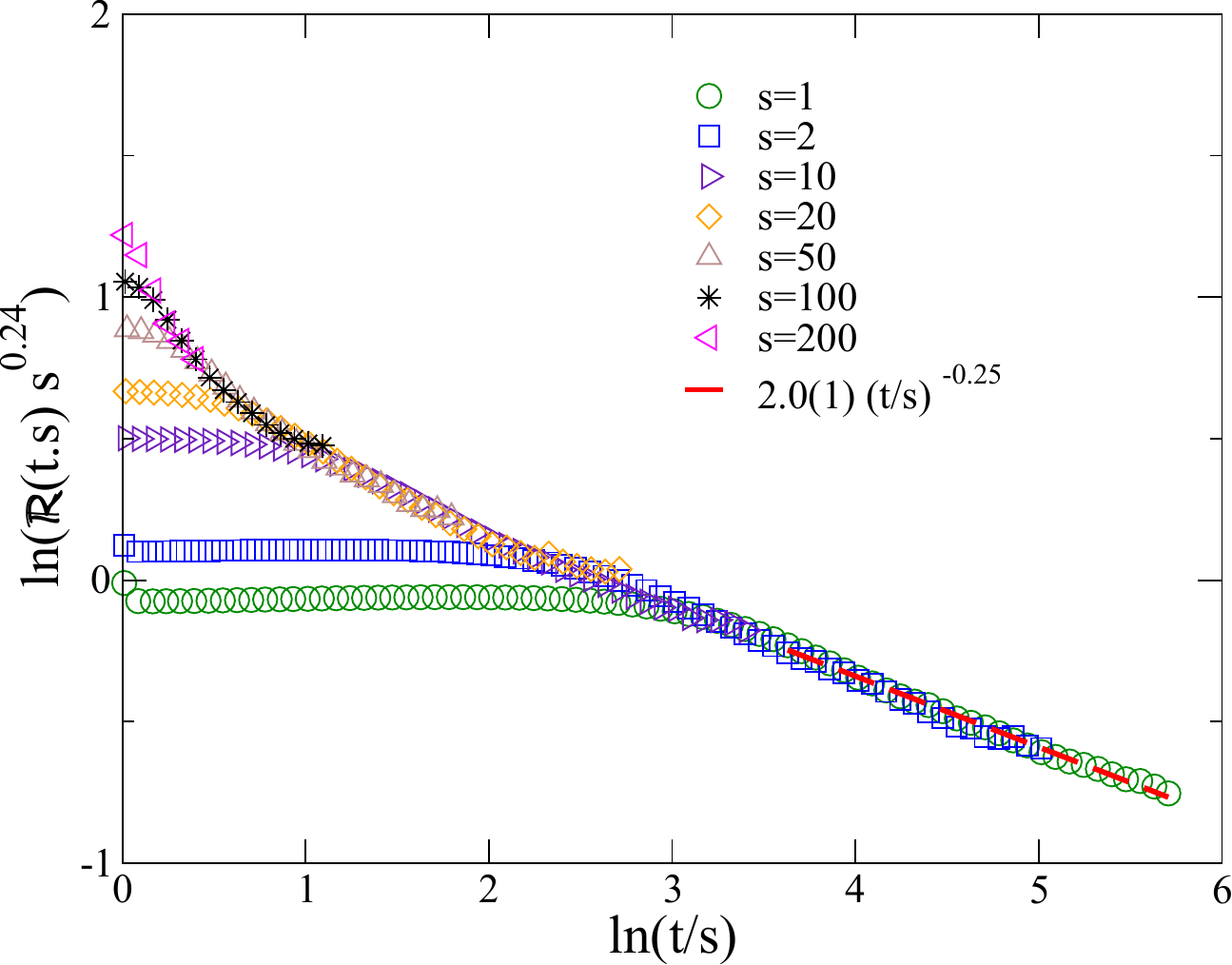}
\caption{Scaling collapse of the auto-responses ${\cal R}(t,s)_w$ in the Kuramoto model for 
different $s$ values indicated by the legends, in case of scenario ({\bf w}) 
at the critical point $K_c=1.9$. The dashed line shows PL fitting for the tail:
$\ln(t/s) > 3.5$. Independent goodness of fit tests are provided in the Appendix II.}
\label{fig:R-Kur-w}
\end{figure}

We applied PL fits for the tails for $\ln(y) > 3$ and the exponents are shown on
the Figures ~\ref{fig:R-Kur-o} (o), \ref{fig:R-Kur-i} (i) and \ref{fig:R-Kur-w} (w).
We have chosen the start of the PL fits for the tails of the data visually, 
which agrees the best the expected straight line of criticality for the largest 
$t/s$ on the log-log plot.
Remarkable data collapse or curves of different $s$ values could also be achieved,
which provides estimates for the short-time aging exponents $1-\text{\textbf{a}}$, appearing to
be close to $\lambda_{\cal R}/Z$ and help to deduce the asymptotic scaling behavior 
for $ y \gg 1$.

\subsubsection{Violation of the fluctuation-dissipation }\label{subsec:vFD}

Now we calculate the FDR via Eq.~(\ref{eq:FDR}) from the numerical results 
of the previous subsections, using the fitted PL tail functions
\begin{equation}
\mathrm{FDR} \simeq X_T(y) \simeq \frac{\mathcal{R}_T(y)} {\dot A_T(y)} \ \ \mathrm{for} \ \ y=t/s\to\infty
\end{equation}
For determining $X_{\infty}$ we assume that at the critical points the FDR denominators
($\dot A_T(y)$) exhibit PL behavior, similarly as the numerators: ${\cal R}(y)$ asymptotically.
The following list summarizes our numerical findings, for FDR, including those for 
the Erd\H os-R\'enyi (ER) random graphs at the Kuramoto model criticality~\cite{Kurcikk}, 
as our reference result. 

\begin{itemize}\centering
   \item $X_T(y)_{(i)} = \displaystyle \frac{\red{3.6(1)}\times y^{-0.35(1)} }{4.0(1) \times 10^{-4} y^{-0.36(3)}} = \red{0.75(2) \times 10^4} y^{0.01(4)} $ 
    \item $X_T(y)_{(o)} \simeq \displaystyle \frac{\red{2.12(1)}\times y^{-0.15(2)}}{7.1(3) \times 10^{-5}y^{-0.14(3)}} = \red{0.30(4) \times 10^4} y^{-0.01(5)}$ 
   \item $X_T(y)_{(w)} = \displaystyle \frac{\red{2.0(1)}\times y^{-0.25(3)} }{14(1)\times 10^{-4} y^{-0.25(3)}} = \red{ 0.14(2) \times 10^4}  y^{0.00(4)}$
   \item $X_T(y)_{(ER)} = \displaystyle \frac{\red{0.23(1)}\times y^{-0.32(2)} }{ 2.2(1)\times 10^{-4} y^{-0.30(2)}} = \red{0.11(2)\times 10^4} y^{-0.02(4)}$
\end{itemize}

As we can see the FDR numerators and denominators exhibit very close PL decay tail exponents.
They completely cancel out for the scenario (w) and their amplitude ratio sequence, 
follows the decreasing trend of the anisotropy values 
($\alpha_i = 0.907$,  $\alpha_o = 0.769$, $\alpha_w = 0.663$, $\alpha_{ER} = 0$),  
i.e. the smaller the anisotropy is, the smaller is the FDR, in the $y\to\infty$ limit, 
summarized in the Table~\ref{tab:sum}.

\begin{table}[h]
\centering
\begin{tabular}{|l|l|l|l|l|l|l|}
\hline
Graph  & $\alpha$ & $X_T\times 10^{-4}$ & $K_c$ & $d_s$ & $\lambda_R/Z$ & $a$ \\
\hline
(i) & 0.907 &  0.75(2) & 0.010(1) & 2.71(1) & 0.36(1) & 0.64(1)\\
(o) & 0.769 &  0.30(2) & 0.0015(5) & 2.107(5) & 0.15(2) & 0.85(2)\\
(w) & 0.663 &  0.14(20) & 1.90(2) & 2.12(2) & 0.25(3) & 0.75(3)\\
ER-d  & 0 &  0.11(2) & 0.482(2) & $\infty$ & 0.32(1) & 0.68(2)\\
ER-o  & 0 &  -- & 0.482(2) & $\infty$ & 0.65(1) & 0.35(5)\\
\hline
\end{tabular}
\caption{Summary of numerical results of for each graph considered. 
The first three rows correspond to versions (scenarios) of the FF.
The KM aging exponents are deduced from the auto-response calculations and
$\lambda_R/Z=1-a$ was applied, which provided good scaling collapses. 
Note, that we have no direct measurements for the aging exponent $b$ and
$\lambda_{\mathcal{R}}=\lambda_A$ seems to emerge from the numerics.
For the ER random graph we have partial/different results for disordered 
(ER-d) and ordered (ER-o) initial conditions, summarized in the table.
Aging exponents in the case of ER are obtained from solutions with
initially random phases, while in the case of FF we could deduce them 
from solutions with phase ordered initial states.}
\label{tab:sum}
\end{table}

The fully reciprocal ER graph exhibits an even smaller FDR amplitude value: 
$X_T \simeq 0.11(1) \times 10^4$ in the large time limit. 
Note, that according to a very recent study~\cite{Henk-age-conf} at criticality 
$\lambda_{\mathcal{R}}=\lambda_A$ is expected, for short-ranged and Gaussian initial correlators, 
which results in simple constant values for the $\lim_{y\to\infty} X(y)$ 
in a good agreement with our numerics.

For the ER model mean-field critical behavior is expected~\cite{Kurcikk}, but 
according to our knowledge no aging exponents are known~\cite{henkel2011non}, 
thus we show our numerical results for a $N=2^{17}$ sized random graph at 
$K_c \simeq 0.482(2)$~\cite{Kurcikk} in the Appendix I. 
As this graph exhibits reciprocity of edges, we can consider it to be the most 
symmetric one of those we investigated here.
Also ER is a graph with long-range interactions. Further study of this random
graph is under way.

Unfortunately we cannot find such relation between anisotropy and FDR in the external driven case, 
because away from criticality both the $\dot A(t)$ and $\mathcal{R}(t)$ saturate to constant values and
their ratio or difference is a constant, that might be related to a temperature of an equilibrium system.

%%%%%%%%%%%%%%%%%%%%%%%%%%%%%%%%%%%%%%%%%%%%%%%%%%%%%%%%%%%%%%%%%%%%%%%%
\section{Conclusions and Outlook}
%%%%%%%%%%%%%%%%%%%%%%%%%%%%%%%%%%%%%%%%%%%%%%%%%%%%%%%%%%%%%%%%%%%%%%%%

In conclusion we have provided numerical evidence for the relation
of vFD and the network asymmetry in agreement with the expectations. 
The higher the link anisotropy is the larger is the FDR, thus the farther 
the system is from equilibrium. 
We determined the auto-response and auto-correlations functions at the 
critical state of the Kuramoto (also for SK) model on the large 
connectome of the fruit-fly as well as for ER graphs. 
As the topological dimension of the network is high, first we expect close to 
mean-field critical behavior, however the spectral dimensions, determining
the synchronization behavior on heterogeneous, non-regular graphs are
found to be smaller than $d_c=5$, thus the temporal behavior of the these
two-point functions is also non-mean-field like, they exhibit slower decay
asymptotically for each anisotropy variant, than in case of the ER network.
Note, however that in these strongly heterogeneous systems Griffiths 
effects~\cite{Griffiths} may also occur by blurring the phase transition 
point as in case of other complex networks~\cite{Atwell01,Frus,Frus-noise,FrusB,llcikk}.
The scaling collapses of the auto-response functions imply the relation:
$1+\text{\textbf{a}}  =  \lambda_{\mathcal{R}/Z}$, very common in aging 
systems~\cite{henkel2011non}. Furthermore, we found numerical support
for the recent claims that $\lambda_{\mathcal{R}}=\lambda_A$, which results in 
simple constant values for the $\lim_{y\to\infty} X(y)$. Our 
results imply that the vFD is related to the opposing deviations of 
the aging behavior, i.e $\dot A(y)$, vs $R(y)$, on the anisotropy 
$\alpha$. 
The simulations also show strong oscillations in the auto-correlations, which is 
typical in systems with non-reciprocal, anisotropic interactions~\cite{PhysRevE.100.062416,evans2026emergentnonreciprocityopenthermodynamicallyconsistent}. 
Furthermore, they can simply be the consequence of the hierarchical modular network structure, which can provide a discrete scale invariance, leading to log-periodic corrections to scaling scaling at criticality~\cite{niemeijer1976renormalization}.\\
Note, that we have neglected the correlator oscillations by using the PL fitted
functions, otherwise using the raw data would have excluded to make any reasonable conclusions.
We think that this kind of data analysis, exploiting scaling assumptions is useful 
and the only possible way to tackle the formidable corrections, coming from oscillations, 
fluctuations and other sources.

The effect of periodic force was shown to decrease the order parameter fluctuations 
as well as sensitivity to $F$.
In terms of the global coupling control parameter $K$, our results
indicate a maximal fluctuations and sensitivity at the critical point: $K_c$.
We also explored the structure of modules of the FF connectome, via community detection 
and by the application of the Cz\'egel-Palla algorithm and we found relatively 
low level of hierarchical organization.

These results provide better understanding of structure and functional relationships in the brain. 
More anisotropic brain parts are farther away from the equilibrium state, while more isotropic ones 
involve closeness to it. This allows identification of regions with higher metabolic rates 
in an alternative way to recent methods~\cite{Lucchetti2025}, or detection of disorders, 
by comparing brains, see for example~\cite{ijms27125190}. 
In the realm of theoretical neuroscience for precision medicine, whole-brain models tailored 
to specific pathologies are increasingly essential for unraveling disease mechanisms and 
improving the predictive accuracy of therapeutic interventions—including pharmacological 
treatments, neurorehabilitation, and brain stimulation. These models enable in silico perturbations, 
offering insights that go beyond static anatomical or functional imaging by probing the brain's 
dynamic capacity and responsiveness.
Recent research has successfully applied FDT-based frameworks to differentiate brain states such as sleep and
wakefulness by quantifying their divergence from FDT-derived expectations~\cite{DecoPhysRevE.108.064410,Monti2025PRR,Patow2024.09.15.613131}. 
This approach offers a robust, theoretically informed lens for interpreting brain dynamics across both healthy and diseased conditions, 
underscoring the relevance and power of FDT in precision neuroscience.
Also vFD can provide a measure for the consciousness level, and test memory effects, that 
can be used for clinical stratification and monitoring, opening avenues for targeted in silico 
intervention planing~\cite{Martinez2025}.
Overall, this framework promises a characterization differentiating brain states in health and disease.

In general, anisotropy, enhances other off-equilibrium signatures and the oscillations.
Note, that we have omitted the effect of time delays, which is in general an important 
factor in brain dynamics~\cite{Petkovski,Cabral}, but we don't expect this to be 
relevant now, as both fluctuation and response effects should propagate in the same way.
Delays have been considered to be irrelevant for the synchronization scaling 
behavior~\cite{Pik}, however a very recent renormalization study suggests, that 
they can induce non-reciprocity~\cite{bae2026kori} in the Kuramoto-Sakaguchi model, 
altering the lower critical dimension leading to non-trivial scaling and interaction 
anisotropy. Further study to verify this would be needed.

As for a future work it remains to be seen the module dependence of the above results 
as well as exploration of the effects of forces on the aging behavior of this complex system.
Moreover, additional, new neurotransmitter data is available now from the 
Flywire collaboration~\cite{down-fullbr}, which could be used to improve the 
investigation of inhibitory effects.

One of the most remarkable findings of our work was the observation of double 
fluctuation peaks in the presence of inhibitory links. This may indicate additional 
dynamical phases or glassy phenomena and a deeper analysis would substantially 
enhance our understanding of the phase structure of the FF as well as other brain systems.
Our method paves the way towards the exploration of the aging behavior
of oscillatory system on different regular or random graphs.

%%%%%%%%%%%%%%%%%%%%%%%%%%%%%%%%%%%%%%%%%%%%%%%%%%%%%%%%%%%%%%%%%%%%%%%%
\section{Acknowledgments}
%%%%%%%%%%%%%%%%%%%%%%%%%%%%%%%%%%%%%%%%%%%%%%%%%%%%%%%%%%%%%%%%%%%%%%%%

We acknowledge support from the National Research, Development and Innovation 
Office NKFIH under Grant No. K146736.  We thank access to the Hungarian national 
supercomputer network via KIF\"U. Initial part of the work was done, in 
2021 using the Marenostrum-4 HPC of the Barcelona Supercomputer Center, 
supported by the HPC-EUROPA3 project. G.\'O. thanks Gustavo Deco and the 
CBC Dep. of the UPF for the hospitality during the visit.
We thank L\'aszl\'o Palla for sharing with us the hierarchy calculating 
Python code, Jeffrey Kelling for maintaining the kuramotoGPU solver and 
Malte Henkel for useful discussions on FDT.

% \clearpage
%%%%%%%%%%%%%%%%%%%%%%%%%%%%%%%%%%%%%%%%%%%%%%%%%%%%%%%%%%%%%%%%%%%%%%%%
\bibliography{bib}
%%%%%%%%%%%%%%%%%%%%%%%%%%%%%%%%%%%%%%%%%%%%%%%%%%%%%%%%%%%%%%%%%%%%%%%%

%\section{Appendix}
\section*{Appendix I. Erd\H os-R\'enyi graph results}

Here we show our aging and FDR results for random ER graphs with average degree 
$\langle k\rangle = 4$.
Most calculations were done for $N=2^{17}$ at $K_c=0.482(1)$, which
was obtained for the Kuramoto model in such networks~\cite{Kurcikk}.
Note, that a that detailed finite size scaling analysis of the
synchronization properties is presented in~\cite{Kurcikk}, leading to the
$K_c=0.482(1)$ estimate and here we also calculated the two-point functions
on smaller sized ER graphs. These results are consistent with the
presented ones, albeit shorter scaling regions could be detected, thus we
just we show the results on the largest sizes we could achieve.

We tried this both for the phase ordered and disordered initial conditions,
but for the ordered (phase-correlated, frequency uncorrelated) case the
${\cal R}(y)$ seems go to a non-zero saturation value for $y\to\infty$ as
some recent works found for the auto-correlator, in case of quenches to the 
ordered phase~\cite{henkelnew}. 
Assuming such saturation value ($ {\cal R}(y\to\infty)=0.02 $), 
we have a rough estimate: $\lambda_{\cal R} / Z = \lambda_A / Z \simeq 0.65(5)$ for  
the ordered initial condition. Furthermore, the $\dot A(y)$ function seems 
to grow with $y$, unlike for the other cases considered here, which provides 
a very uncertain estimate for the FDR. Further investigation of this is
under way and will be published somewhere else.

\begin{figure}[h]\centering
\includegraphics[width=0.7\columnwidth]{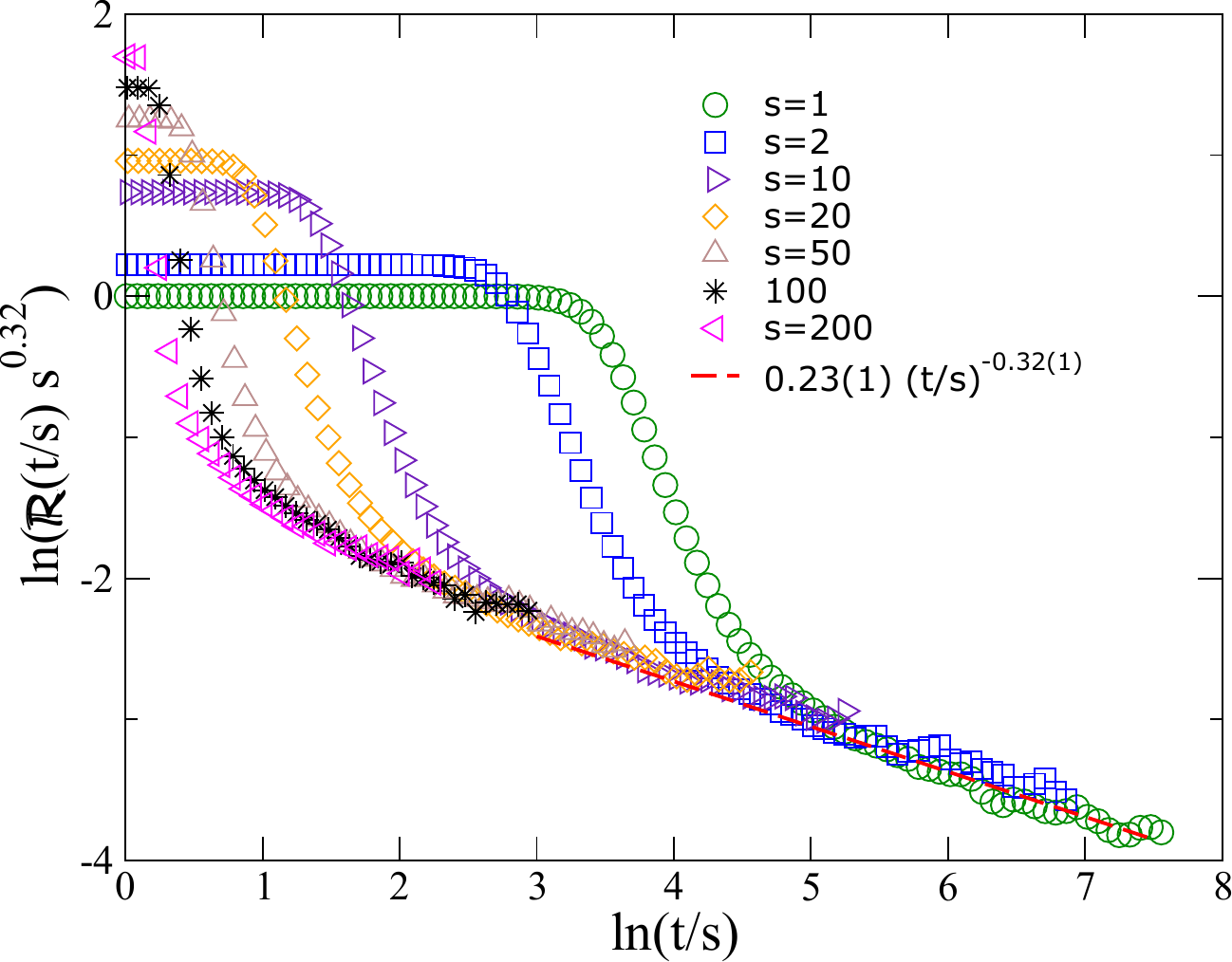}
\caption{Scaling collapse of the auto-responses in the Kuramoto model for 
$s$ values indicated by the legends in case of the ER graph at the 
critical point $K_c\simeq 0.482$. These solutions correspond to 
uniformly distributed random initial phases.
The dashed line shows a least squares PL fitting for the tail: 
$\ln(t/s) > 3$, with an exponent: $\lambda_R/Z=0.32(2)$. 
Independent goodness of fit tests are provided in the Appendix II.}
\label{fig:R-ER-TM2}
\end{figure}

\begin{figure}[h]\centering
\includegraphics[width=0.75\columnwidth]{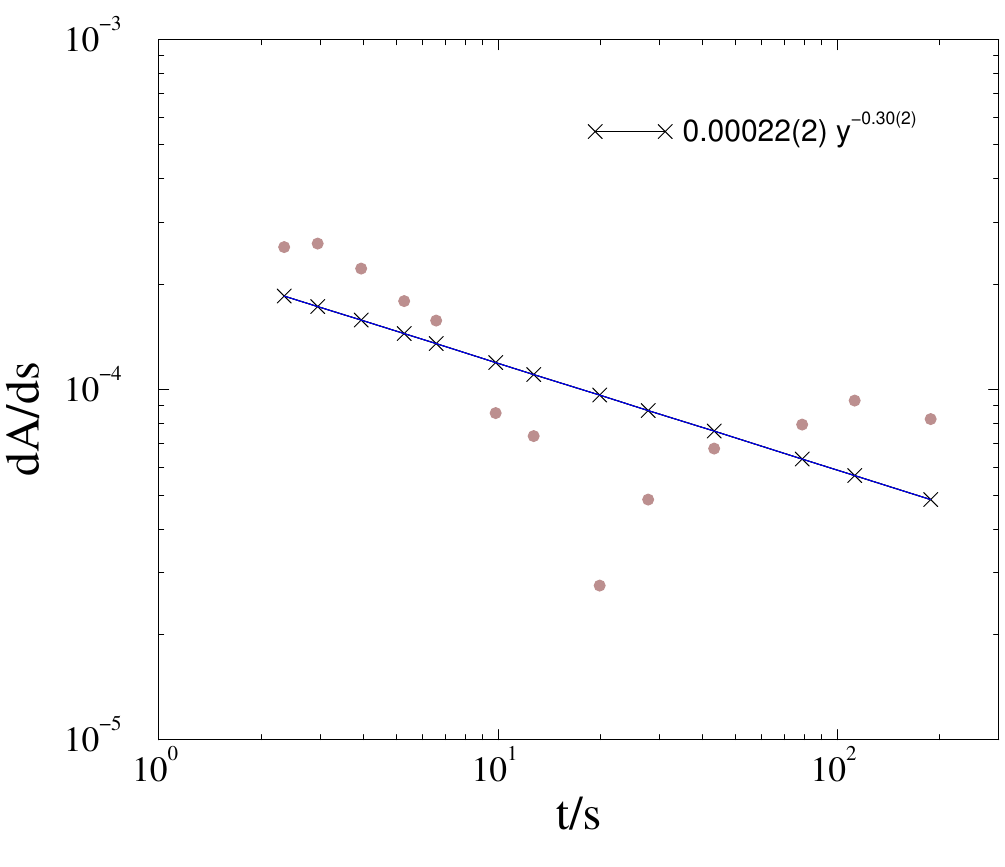}
\caption{Here we see the $s$ derivatives of the auto-correlations: $\dot A(t,s)$
of the critical Kuramoto on ER graph in case of phase disordered initial state,
deduced from the results of $t=120$ and $258$ as shown by the bullets.
The line shows a least squares PL fitting assuming with the exponent $-0.30(2)$.
Independent goodness of fit tests are provided in the Appendix II.}
\label{fig:dA-ERc-TM2}
\end{figure}

In the case of random (fully uncorrelated) phases and self-frequencies 
we obtained a reasonable scaling within or limitations of computing 
possibilities $\lambda_{\cal R} / Z = \lambda_A / Z \simeq 0.32(5)$ 
as shown in Figs.~\ref{fig:R-ER-TM2} and~\ref{fig:dA-ERc-TM2}. 
Here averaging was performed for $\simeq 600$ realizations up to $t_{max}=2000$.
As the fitted tail exponents of $R(y)$ and $\dot A(y)$ are very close in the 
FDR ratio, they almost cancel out, suggesting closeness to the equilibrium 
in case of this reciprocal and symmetric graph with the Kuramoto dynamics.

\section*{Appendix II. Fitting validation by other tests}

To validate the fitting parameters additional statistics were calculated, we performed the  non-parametric Kolmogorov–Smirnov (KS) test and the  Akaike Information Criterion (AIC) test. 
The KS is used to determine if a sample follows a specific distribution or if two independent samples are subsets of the same distribution. The core concept of the KS is working by comparing Cumulative Distribution Functions (CDFs). We performed the "D" test variation, which also calculates a value \(D\), which is the maximum absolute difference between the two CDFs being compared. The smaller the D value the closer the data is to the distribution. The KS p-value tells the statistical significance of distance \(D\). It helps deciding whether to "reject" the assumption that the data follows the required distribution. This value is between 0 and 1. If this value is below 0.05, then the two are from two different distributions, rejection can be considered. We show in table \ref{tab:fit-metrics-summary} that our fits performed very well in most cases giving close 
to 1 $p$ values.

Furthermore, we evaluate the Akaike Information Criterion with least square model fitting: $\text{AIC} = N \ln(\text{RSS}/N) + 2k$, where $k$ is the number of fit parameters, $N$ is the number of tail data points and $RSS$ is the residual sum of squares or sum of squared errors ($RSS = \sum_{i=1}^{n} (y_i - \hat{y}_i)^2$, here $y_i$ are the empirical values in the tail and $\hat{y}_i$ are the points predicted by the model). Normally when comparing models we would assume $\text{AIC} = 2k-2\ln(L$), where $L$ is the the maximum likelihood of the model on the data. However, for small sample size there is a substabtial probability that the AIC will select models with too many parameters, 
We compare the power-law model against exponential (EXP) and log-normal (LN) models over the same scaling window by computing the relative AIC: $\Delta\text{AIC}_{\text{EXP}}$ and $\Delta\text{AIC}_{\text{LN}}$, where $\Delta\text{AIC}_{\text{EXP/LN}} = \text{AIC}_{\text{PL}} - \text{AIC}_{\text{EXP/LN}}$. The more negative $\langle \Delta\text{AIC} \rangle$ values indicate that the power-law model is preferred because it yields a lower information loss compared to the alternative distributions. The resulting $\langle \Delta\text{AIC} \rangle$ statistics robustly confirm the high fidelity of the power-law fits across all network scenarios.
\begin{table}[h]
\centering
\begin{tabular}{|l|l|l|l|l|l|}
\hline
Figure & Scenario & D value & KS p-value & $\Delta$AIC (EXP) & $\Delta$AIC (LN) \\
\hline
Figure-8 & o & 0.090 & 0.558 & -243.335 & 36.390 \\
         & i & 0.048 & 1.000 & -285.257 & -1.955 \\
         & w & 0.067 & 0.974 & -256.056 & 24.250 \\
Figure-9 & o & 0.154 & 0.999 & -3.695 & -1.402 \\
         & i & 0.375 & 0.660 & -6.970 & 4.474 \\
         & w & 0.538 & 0.044 & 0.159 & -1.607 \\
Figure-10 & o & 0.118 & 0.738 & -93.078 & -81.318 \\
Figure-11 & i & 0.028 & 1.000 & -179.195 & 7.731 \\
Figure-12 & w & 0.036 & 1.000 & -89.642 & -49.478 \\
Figure-13 & ER & 0.129 & 0.963 & -49.823 & 19.259 \\
\hline
\end{tabular}
\centering
\caption{Goodness-of-fit and model selection summary across scaling scenarios: comparison of power-law fit quality across figures and network scenarios ((o): original network, (i): inhibition model, (w): randomized weights, $ER$: Erd\H{o}s--R\'enyi graph). The Kolmogorov--Smirnov (KS) statistic $D$ and high $p$-values ($p > 0.05$) confirm that the power-law hypothesis cannot be rejected in most cases. Akaike Information Criterion ($\text{AIC}$) statistics evaluate model performance.}

\label{tab:fit-metrics-summary}
\end{table}

\end{document}